\documentclass[12pt]{article}
\pdfoutput=1
\usepackage{geometry,amsmath,amsfonts}
\usepackage{slashed}
\usepackage{epsfig}
\usepackage{latexsym}
\usepackage{graphicx}
\usepackage[justification=RaggedRight]{caption}
\usepackage{amssymb}
\usepackage{subfig}
\usepackage{color}
\usepackage{multirow}
\usepackage{color}
\usepackage{rotating}
\usepackage{ifthen}
\usepackage{epsfig}
\usepackage{hyperref}
\usepackage{cite}

\newcommand{\gsim}{\lower.7ex\hbox{$\;\stackrel{\textstyle>}{\sim}\;$}}
\newcommand{\lsim}{\lower.7ex\hbox{$\;\stackrel{\textstyle<}{\sim}\;$}}

\begin{document}

\vspace*{5mm} 

\begin{center}

\mbox{\large\bf 
A possible interpretation of the LUX-ZEPLIN recoil event in}  

\vspace*{2mm}

\mbox{\large\bf the 2HD+a scenario}

\vspace*{5mm}

{\sc Giorgio Arcadi}$^{1,2}$, {\sc Mattia di Mauro}$^{3}$, {\sc Abdelhak~Djouadi}$^{4}$ \\ and {\sc Farinaldo Queiroz}$^{5,6,7}$ 

\vspace{5mm}

{\small 

$^1$ Dipartimento di Scienze Matematiche e Informatiche, Scienze Fisiche e Scienze della Terra, \\
Universita degli Studi di Messina, V. Ferdinando Stagno d'Alcontres 31, I-98166 Messina, Italy.\\[2mm]

$^2$ INFN Sezione di Catania, Via Santa Sofia 64, I-95123 Catania, Italy. \\[2mm]

\mbox{ \hspace*{-6mm} $^3$ INFN Sezione di Torino, Via Pietro Giuria 1, 10125 Torino, Italy} \\[2mm] 

\mbox{ \hspace*{-6mm} $^4$ Departamento de F\'isica Te\'orica y del Cosmos, Universidad de Granada,
18071 Granada, Spain.} \\[2mm] 

 \hspace*{-6mm}
$^5$ International Institute of Physics, Universidade Federal do Rio Grande do Norte, Campus Universitário, Lagoa Nova, Natal-RN 59078-970, Brazil. \\ \vspace{0.2cm}

 \hspace*{-6mm}
$^6$ Departamento de F\'isica, Facultad de Ciencias, Universidad de La Serena,
Avenida Cisternas 1200, La Serena, Chile  \\ \vspace{0.2cm}

 \hspace*{-6mm}
$^7$  Millennium Institute for Subatomic Physics at High-Energy Frontier (SAPHIR),
Fernandez Concha 700, Santiago, Chile. \\ \vspace{0.2cm}.

}

\end{center}

\vspace*{1cm}










\begin{abstract} 

It is tantalizing to consider the recent observation of the LUX-ZEPLIN (LZ) experiment of a high energy nuclear recoil candidate at $E_{\rm Recoil} \simeq 248$ keV as the long-awaited signal of a weakly interacting massive  Dark Matter (DM) particle. Although this signal is still weak and needs to be confirmed by further data and inspection, we attempt to interpret it in the context of the 2HD+a model in which a two-Higgs doublet model (2HDM) is supplemented by a light pseudoscalar Higgs boson $a$ and an isosinglet fermionic DM particle $\chi$. The model has the virtue of reproducing the correct cosmological density and evading the previous limits from direct and indirect detection of DM particles,  while passing all constraints from colliders searches and high-precision measurements in the Higgs, electroweak boson and heavy-flavor sectors. We indeed find regions of the parameter space of the model, with a rather heavy 2HDM spectrum but a very light pseudoscalar $a$ boson, $m_a \approx 1$--10 GeV, as well as a DM fermion with a mass of a few hundred GeV,  which are compatible with the LZ recoil event. The related extensive investigation of the parameter space of the 2HD+a model with fermionic DM, including the wide range of complementary constraints from both DM phenomenology and collider searches as well as precision measurements are presented. 
\end{abstract}


\newpage

\section{Introduction}

The nature of dark matter (DM) remains one of the main open questions in particle physics and cosmology. Among the many proposed particle candidates, weakly interacting massive particles have been extensively studied because their interactions with the Standard Model (SM) can simultaneously determine the cosmological abundance through thermal freeze-out and produce observable signatures in direct-detection, indirect-detection, and collider experiments \cite{Arcadi:2019lka,Arcadi:2024ukq}. At the same time, the increasingly stringent constraints from direct-detection experiments have excluded large regions of the parameter space associated with conventional spin-independent (SI) interactions \cite{Arcadi2018,Arcadi:2019lka,DiMauroArina2023,LZ2025WIMP,XENONnT2025,DiMauroXie2026,KongDiMauro2026,KoechlerDiMauro2025}, motivating scenarios characterized by non-standard nuclear responses, momentum-dependent interactions, or multiple dark-sector states \cite{PospelovRitzVoloshin2008,ArcadiMambriniPierre2015,ArcadiBenincasaDjouadiKannike2022,DiMauroArina2023,DiMauroXie2025,DiMauroWang2025,ShaikhDiMauro2026}.

Recently, the LUX-ZEPLIN Collaboration (LZ) extended its nuclear-recoil analysis to energies substantially above those usually considered in standard DM searches \cite{LZ:2026extended}. Using an exposure of $2.84~{\rm tonne,yr}$ and a recoil-energy window extending up to approximately $270~{\rm keV}$, LZ reported a single event compatible with a nuclear recoil of
\begin{equation}
E_R = 248 \pm 23_{\rm stat} \pm 23_{\rm sys}~{\rm keV}.
\end{equation}
The event lies in a region where the expected background is very small. Depending on the assumed DM interaction, the largest local significance reaches $3.4\sigma$, and is reduced to $2.6\sigma$ after accounting for the look-elsewhere effect \cite{LZ:2026extended}. Although the statistical significance is not sufficient to claim evidence for DM, the unusually large recoil energy, together with the absence of a corresponding excess at lower recoil energies, makes this event particularly interesting from a phenomenological perspective.

A significant number of interpretations have appeared following the LZ result. A broad class of explanations relies on endothermic inelastic DM, in which the scattering process $\chi_1 N \rightarrow \chi_2 N$ produces a heavier dark-sector state. A mass splitting of order a few hundred keV suppresses scattering at low recoil energies and shifts the spectrum toward the high-energy region \cite{Su:2026iDM,DiMauro:2026Kinematic,McCabe:2026Seasonal}. Particularly predictive examples include quasi-Dirac Higgsino DM \cite{Freese:2026Higgsino,DiMauro:2026Kinematic}, together with ultraviolet realizations involving high-scale supersymmetry, non-universal gaugino masses, singlet-doublet fermions, and extended neutralino sectors \cite{Yin:2026PQSUSY,DuWang:2026HiggsinoGauginos,Bisal:2026GNMSSM,Borah:2026SingletDoublet,LeeYoun:2026Mixing}. Inelastic scalar realizations have also been investigated in inert-doublet and extended gauge models \cite{Nomura:2026Z2Partner,WangXiao:2026IDM,OkadaSeto:2026BL,KumarPrajapati:2026ChiralBL}.

These interpretations can be subject to important complementary constraints. In particular, DM particles can reach substantially larger velocities in the Solar interior than in terrestrial detectors. Consequently, inelastic channels that are close to their kinematic threshold on Earth can remain open inside the Sun. Solar capture followed by DM annihilation can then produce high-energy neutrinos, leading to strong IceCube constraints on some LZ-motivated models, most notably the canonical thermal Higgsino interpretation \cite{PospelovRamani:2026Solar,DiMauroShaikh:2026Solar,NguyenLindenHooper:2026Solar}. Alternative possibilities include heavier or non-standard Higgsino scenarios \cite{Langhoff:2026HeavyHiggsino}. Other proposed explanations avoid standard endothermic halo scattering altogether. These include exothermic DM \cite{BaerBarger:2026Exothermic,Fan:2026iDM}, boosted dark-sector particles \cite{Alhazmi:2026Boosted,Liang:2026BoostedDipole,Kannike:2026Boosted}, fermionic DM absorption \cite{LouLu:2026Absorption}, and neutron-disappearance processes in xenon \cite{AghaieStrumia:2026Neutron}. Taken together, these studies illustrate that the LZ event can probe a broad range of dark-sector kinematics and interactions.

A qualitatively different possibility is that the event originates from elastic DM scattering with a strongly momentum-dependent interaction. Indeed, the LZ analysis did not restrict its interpretation to inelastic scattering, but also considered a broad set of elastic effective interactions. In particular, in an effective field theory (EFT) approach, the conventional spin-dependent (SD) operator $\mathcal{O}_4$ reaches a local significance of approximately $2.7\sigma$ for DM mass of $m_\chi = 1~{\rm TeV}$. Several momentum-dependent elastic interactions provide an even better description of the high-energy recoil event. For example, again in the EFT approach, the covariant interaction $\mathcal{L}_{10}^{N}$ contains, in the non-relativistic limit, momentum-dependent contributions involving the operators $\mathcal{O}_4$ and $\mathcal{O}_6$. For an isoscalar coupling, $\mathcal{L}_{10}^{s}$ reaches a local significance of $3.4\sigma$ at $m_\chi = 1~{\rm TeV}$, while the isoscalar interaction $\mathcal{L}_{16}^{s}$ reaches the same local significance. Therefore, the largest local preference found by LZ is not exclusive to inelastic scattering. Elastic interactions characterized by non-standard momentum dependence and nuclear responses can provide an equally good description of the $248~{\rm keV}$ recoil event.

The difficulty of explaining such a high-energy recoil through a conventional SI interaction is not purely kinematic. Standard SI scattering generally predicts a recoil spectrum strongly weighted toward low energies, while the progressive loss of nuclear coherence at large momentum transfer further suppresses the high-energy tail. Momentum-dependent and SD interactions can behave very differently. Additional powers of the momentum transfer, together with non-trivial nuclear response functions, can substantially modify the recoil spectrum and enhance the relative importance of large recoil energies. Elastic pseudoscalar interactions have already been discussed in this context in axion-portal constructions \cite{Unwin:2026AxionPortal}, while non-trivial nuclear-response effects can generate characteristic spectral structures that differ significantly from those expected for conventional SI scattering \cite{Khan:2026Nuclear}.

In this work, we investigate this elastic interpretation within the two-Higgs-doublet model (2HDM) supplemented by a pseudoscalar singlet $a$ and a fermionic DM particle $\chi$, hereafter referred to as the 2HD+a model \cite{Ipek:2014gua,Bauer:2017ota,Arcadi:2020gge,Arcadi:2022lpp}. The model provides a gauge-invariant and renormalizable realization of pseudoscalar interactions between DM and SM fermions. Mixing between the pseudoscalar singlet and the CP-odd state of the 2HDM sector allows the light physical pseudoscalar $a$ to couple simultaneously to the dark and visible sectors. In the non-relativistic limit, the resulting DM--nucleon interaction is dominated by the momentum-suppressed operator $\mathcal{O}_6^{\rm NR}$. The corresponding strong momentum dependence makes the model particularly well suited to generating a nuclear-recoil spectrum with an enhanced high-energy component.

A particularly relevant feature of the 2HD+a framework is that the parameters controlling the LZ signal also determine several complementary observables. The couplings of the pseudoscalar mediator to SM fermions depend on the Yukawa realization of the underlying 2HDM, giving rise to four different types of scenarios. At the same time, the DM Yukawa coupling and the pseudoscalar mixing angle control the DM annihilation rate and thus the thermal relic abundance. The same parameter space is further constrained by loop-induced SI scattering, flavor observables, high precision measurements and direct collider Higgs searches.

We calculate the full momentum-dependent elastic scattering rate in xenon, including the interference between the light and heavy pseudoscalar propagators and the relevant nuclear response functions. We determine the values of $y_\chi$ required to reproduce the LZ event as functions of $m_\chi$, $m_a$, $\tan\beta$, and the pseudoscalar mixing angle, considering all four Yukawa realizations. We then confront the LZ-favored regions with constraints from loop-induced SI scattering, collider searches, and flavor observables, with particular emphasis on the exotic Higgs decay $h \rightarrow aa$ and searches for light pseudoscalars. Finally, we impose the thermal relic-density requirement and investigate whether the same parameter configurations can simultaneously reproduce the LZ event and the observed DM abundance.

The central question addressed in this work is therefore whether the high-energy LZ recoil can be explained without introducing an inelastic dark-sector mass splitting, relying instead on the intrinsic momentum dependence of pseudoscalar-mediated elastic scattering within a complete and phenomenologically constrained particle-physics model.

The paper is organized as follows. In Sec.~\ref{sec:model} we introduce the 2HD+a model and its four Yukawa types or  realizations. In Sec.~\ref{sec:dd} we discuss the elastic direct-detection signal and determine the parameter space selected by the LZ event. In Sec.~\ref{sec:collider} we analyze the relevant collider, flavor, and loop-induced direct-detection constraints. In Sec.~\ref{sec:relic} we impose the thermal relic-density requirement and in Sec.~\ref{sec:combined} we combined the existing collider and relic density bounds with the LZ preferred parameter space. Finally, we summarize our conclusions in Sec.~\ref{sec:conclusions}.

\section{The 2HD+a model}
\label{sec:model}

The 2HD+a model extends the scalar sector of the SM by introducing a second Higgs doublet and a real CP-odd gauge singlet. The scalar sector is therefore composed of two ${\rm SU(2)_L}$ doublets, $\Phi_1$ and $\Phi_2$, together with the pseudoscalar singlet $a_0$. The most general CP-conserving scalar potential relevant for our analysis can be written as \cite{Bauer:2017ota,Ipek:2014gua,Arcadi:2022dmt,Arcadi:2022lpp,Robens:2021lov,Arcadi:2020gge,Argyropoulos:2022ezr,Argyropoulos:2024yxo}
\begin{eqnarray}
V = V(\Phi_1,\Phi_2)
+\frac{1}{2}m_{a_0}^2 a_0^2
+\frac{\lambda_a}{4}a_0^4
+\left(i\kappa a_0\Phi_1^\dagger\Phi_2+\mbox{h.c.}\right)
+\lambda_{1P}a_0^2\Phi_1^\dagger\Phi_1
+\lambda_{2P}a_0^2\Phi_2^\dagger\Phi_2 \, ,
\end{eqnarray}
where $V(\Phi_1,\Phi_2)$ denotes the usual CP-conserving two-Higgs-doublet-model (2HDM) potential \cite{Branco:2011iw}. 

After electroweak symmetry breaking, the neutral components of the two doublets acquire vacuum expectation values $v_1$ and $v_2$, which we parametrize as
\begin{equation}
v_1=v\cos\beta,
\qquad
v_2=v\sin\beta,
\qquad
\tan\beta=\frac{v_2}{v_1},
\qquad
v=\sqrt{v_1^2+v_2^2}\simeq246~{\rm GeV}.
\end{equation}
The physical scalar spectrum then consists of two neutral CP-even states, $h$ and $H$, a charged scalar pair $H^\pm$, and two neutral CP-odd states, $a$ and $A$. We identify the lighter CP-even state $h$ with the observed Higgs boson with mass $M_h\simeq125~{\rm GeV}$, while $H$ denotes the additional heavy CP-even state. The angle $\alpha$ parametrizes the mixing in the CP-even sector.

The two physical CP-odd states arise from the mixing between the doublet pseudoscalar $A_0$ and the singlet state $a_0$. We define the mass eigenstates through
\begin{equation}
\left(
\begin{array}{c}
A_0\\
a_0
\end{array}
\right)
=
\left(
\begin{array}{cc}
\cos\theta & \sin\theta\\
-\sin\theta & \cos\theta
\end{array}
\right)
\left(
\begin{array}{c}
A\\
a
\end{array}
\right),
\qquad
\tan 2\theta=
\frac{2\kappa v}{M_A^2-m_a^2}.
\end{equation}
Unless otherwise stated, we take $m_a<M_A$, so that $a$ denotes the lighter pseudoscalar mediator relevant for the DM phenomenology discussed below.
In the mass basis, the scalar sector can be parametrized in terms of the five physical masses $M_h$, $M_H$, $M_{H^\pm}$, $M_A$, and $m_a$, the three mixing angles $\alpha$, $\beta$, and $\theta$, together with three independent couplings of the scalar potential, which we choose as $\lambda_{1P}$, $\lambda_{2P}$, and $\lambda_3$.

A central feature of the 2HD+a construction is that the mixing between $A_0$ and $a_0$ generates a gauge-invariant coupling of the singlet-like pseudoscalar to SM fermions. The Yukawa interactions of the neutral scalar states with SM fermions can be written as
\begin{eqnarray}
\mathcal{L}_{\rm Yukawa}
=
\sum_f\frac{m_f}{v}
\left[
g_{hff}\,h\bar f f
+
g_{Hff}\,H\bar f f
-
i g_{A_0ff}\cos\theta\,A\bar f\gamma_5 f
+
i g_{A_0ff}\sin\theta\,a\bar f\gamma_5 f
\right].
\end{eqnarray}

The $H^\pm$ couplings follow from the same Yukawa structure of the underlying 2HDM.

The DM candidate is taken to be a SM-singlet fermion $\chi$, which couples directly to the pseudoscalar singlet component. In the physical basis, the corresponding interaction is
\begin{equation}
\mathcal{L}_{\chi}
=
i y_\chi
\bar\chi\gamma_5\chi
\left(
a\cos\theta+A\sin\theta
\right),
\end{equation}
where $y_\chi$ denotes the DM Yukawa coupling. The light pseudoscalar $a$ therefore couples to the DM sector through its singlet component, proportional to $\cos\theta$, and to SM fermions through its doublet component, proportional to $\sin\theta$. This structure is particularly important for the direct-detection phenomenology, since the tree-level DM--nucleon amplitude is proportional to the product $y_\chi\sin\theta\cos\theta$.

The absence of tree-level FCNCs restricts the Yukawa sector to the usual four 2HDM realizations \cite{Branco:2011iw}, conventionally referred to as Type-I, Type-II, Type-X, and Type-Y. These four possibilities correspond to different assignments of the two Higgs doublets to the SM fermions. In Type-I, all quarks and charged leptons couple to the same doublet, conventionally chosen as $\Phi_2$. In Type-II, up-type quarks couple to $\Phi_2$, while down-type quarks and charged leptons couple to $\Phi_1$. In Type-X, also referred to as the lepton-specific scenario, all quarks couple to $\Phi_2$, whereas charged leptons couple to $\Phi_1$. Finally, in Type-Y, also known as the flipped scenario, up-type quarks and charged leptons couple to $\Phi_2$, while down-type quarks couple to $\Phi_1$. These different assignments lead to characteristic dependences of the fermionic couplings on $\tan\beta$ and therefore to substantially different collider, flavor, and DM phenomenology.
Table~\ref{table:2hdm_type} summarizes the corresponding normalized couplings of the additional neutral Higgs states to up-type quarks, down-type quarks, and charged leptons for the four Yukawa realizations in the alignment limit.

Throughout this work, we assume the alignment limit,
\begin{equation}
\beta-\alpha=\frac{\pi}{2},
\end{equation}
in which the light CP-even state $h$ has SM-like couplings, $g_{hff}=1$. In this limit, the couplings of the additional neutral scalar states to fermions can be expressed as
\begin{equation}
g_{Hff}=g_{\phi ff},
\qquad
g_{Aff}=\cos\theta\,g_{\phi ff},
\qquad
g_{aff}=-\sin\theta\,g_{\phi ff},
\end{equation}
where the coefficients $g_{\phi ff}$ depend only on $\tan\beta$ and on the Yukawa realization. 
Their explicit values are summarized in Table~\ref{table:2hdm_type}.

\begin{table}[h!]
\renewcommand{\arraystretch}{1.4}
\begin{center}
\begin{tabular}{|c|c|c|c|c|}
\hline
~~~~~~~~~~~~ & ~~~Type-I~~~ & ~~~Type-II~~~ & ~~~Type-X~~~ & ~~~Type-Y~~~ \\ \hline
$g_{\phi tt}$ & $\frac{1}{\tan\beta}$ & $\frac{1}{\tan\beta}$ & $\frac{1}{\tan\beta}$ & $\frac{1}{\tan\beta}$ \\ \hline
$g_{\phi bb}$ & $-\frac{1}{\tan\beta}$ & $\tan\beta$ & $-\frac{1}{\tan\beta}$ & $\tan\beta$ \\ \hline
$g_{\phi\tau\tau}$ & $-\frac{1}{\tan\beta}$ & $\tan\beta$ & $\tan\beta$ & $-\frac{1}{\tan\beta}$ \\ \hline
\end{tabular}
\vspace*{-.1mm}
\caption{Couplings of the additional neutral doublet-like Higgs states to SM fermions, normalized to the corresponding SM Yukawa coupling, for the four Yukawa realizations of the 2HDM. The expressions are given in the alignment limit, $\beta-\alpha=\pi/2$, for which the light CP-even state has SM-like couplings, $g_{htt}=g_{hbb}=g_{h\tau\tau}=1$.}
\label{table:2hdm_type}
\end{center}
\vspace*{-9mm}
\end{table}

The different $\tan\beta$ dependence of these couplings will play an important role in the phenomenology of the LZ event. In all four realizations the top-quark coupling scales as $g_{\phi tt}=\cot\beta$, so very small values of $\tan\beta$ lead to a rapidly increasing top Yukawa coupling. Requiring perturbative couplings therefore motivates approximately $\tan\beta\gsim1/3$. At large $\tan\beta$, the relevant constraint depends on the Yukawa realization. In Type-X, the $\tau$ coupling grows as $\tan\beta$, suggesting $\tan\beta\lsim{\cal O}(100)$, whereas in Type-II and Type-Y the bottom-quark coupling is enhanced and perturbativity typically requires $\tan\beta\lsim50$--60. If leptonic couplings are neglected, Type-I and Type-X have identical quark Yukawa structures, as do Type-II and Type-Y. This observation will be useful when discussing the DM--nucleus scattering rate.

The model is also subject to several theoretical consistency conditions. As in a conventional 2HDM, the scalar potential must be bounded from below, ensuring that it does not develop runaway directions toward arbitrarily negative values of the energy and that a stable electroweak vacuum can exist. In addition, the quartic couplings must satisfy perturbative unitarity constraints, which require the eigenvalues of the relevant scalar and electroweak gauge-boson scattering amplitudes to remain sufficiently small for the perturbative description of the theory to be reliable. These conditions restrict the allowed combinations of quartic couplings and, indirectly, the possible mass splittings among the additional Higgs states. The corresponding constraints have been studied in detail, for example, in Ref.~\cite{Arcadi:2022lpp}, and are included in our numerical analysis.


A second class of constraints arises from collider, flavor, and electroweak precision measurements \cite{Branco:2011iw,Arcadi:2022lpp,Argyropoulos:2022ezr,Argyropoulos:2024yxo}. Measurements of the properties of the $125~{\rm GeV}$ Higgs boson strongly favor the alignment regime adopted above, while direct searches constrain the additional states $H$, $A$, and $H^\pm$. Further restrictions arise from electroweak precision observables, which constrain the allowed masses and mass splittings within the extended scalar sector, and from flavor-changing processes involving the additional Higgs states \cite{Branco:2011iw,Arcadi:2022lpp}.

A particularly important example comes from $b\rightarrow s$ transitions. In Type-II and Type-Y scenarios, the experimental world average of the inclusive branching ratio ${\rm BR}(B\to X_s\gamma)$, obtained by HFLAV from measurements by BaBar, Belle, and CLEO, provides a particularly strong constraint on the charged-Higgs sector. Comparing this measurement with the NNLO theoretical prediction yields approximately
\begin{equation}
M_{H^\pm}\gtrsim 800~{\rm GeV},
\end{equation}
for the Type-II and Type-Y Yukawa realizations \cite{Misiak:2020vlo}.
Electroweak precision observables and theoretical consistency conditions generally disfavor large mass splittings among the heavy scalar states. We therefore adopt as a representative configuration a heavy and approximately degenerate 2HDM spectrum,
\begin{equation}
M_A\simeq M_H\simeq M_{H^\pm}\gtrsim800~{\rm GeV},
\end{equation}
together with the alignment limit. This choice efficiently suppresses several constraints associated with the heavy Higgs sector while allowing us to focus on the light pseudoscalar $a$, which plays a central role in the interpretation of the LZ event.

The phenomenology of a light pseudoscalar is itself subject to important experimental constraints. For the mass range considered, three classes of observables are particularly relevant. First, the pseudoscalar contributes to flavor observables such as $B_s\to\mu^+\mu^-$ and to the anomalous magnetic moment of the muon, $(g-2)_\mu$, with a sensitivity that depends strongly on its coupling to leptons \cite{Arcadi:2022lpp,Arcadi:2022dmt,Arcadi:2026aau}. Second, direct searches for low-mass dimuon resonances constrain the process $pp\to a\to\mu^+\mu^-$, which has been investigated by ATLAS, CMS, and LHCb \cite{ATLAS:2026bpi,CMS:2023hwl,LHCb:2020ysn}. These bounds become particularly important in Yukawa realizations in which the coupling of $a$ to leptons or down-type quarks is enhanced.

Finally, for $m_a<\frac12 M_h$, the exotic decay $h\to aa$ is kinematically open. Since the observed Higgs boson is experimentally constrained to remain close to its SM behavior, the corresponding trilinear coupling $haa$ must be sufficiently suppressed \cite{Arcadi:2022lpp}. As discussed in Sec.~\ref{sec:collider}, satisfying this requirement can impose non-trivial relations among the parameters of the scalar potential.

These theoretical, collider, and flavor constraints define the general viable parameter space of the 2HD+a model. Additional restrictions are directly associated with the DM sector and with the interpretation of the LZ recoil event. In particular, the parameters $y_\chi$, $\theta$, $m_a$, $\tan\beta$, and $m_\chi$ simultaneously determine the momentum-dependent elastic scattering rate, the loop-induced SI interaction, and the thermal annihilation rate. We discuss these complementary requirements in the following sections, beginning with the direct-detection signal.

\section{Direct Detection in the 2HD+a Model}
\label{sec:dd}

\subsection{Tree-level direct detection and fit of the LZ excess}
\label{sec:dd_tree}

We now discuss the direct-detection phenomenology of the 2HD+a model and determine the regions of parameter space that can reproduce the high-energy LZ recoil event. At tree level, elastic DM scattering off nuclei is mediated by the exchange of the two pseudoscalar mass eigenstates $a$ and $A$. Their Yukawa couplings to SM quarks and to the DM fermion generate the effective quark-level interaction,
\begin{equation}
c_q(Q^2)\,\bar{\chi}\gamma_5\chi\,\bar{q}\gamma_5 q,
\end{equation}
with
\begin{equation}
c_q(Q^2)
=
y_\chi \sin\theta\cos\theta\,
\frac{m_q}{v}\,
g_{Aqq}
\left[
\frac{1}{Q^2+m_a^2}
-
\frac{1}{Q^2+M_A^2}
\right].
\label{eq:cq_dd}
\end{equation}
Here $\mathbf{Q}$ denotes the three-momentum transferred to the target nucleus and $Q^2\equiv |\mathbf{Q}|^2$. For a nuclear target of mass $m_T$ and recoil energy $E_R$, the momentum transfer is related to the recoil energy through
\begin{equation}
Q^2=2m_T E_R.
\label{eq:q_er}
\end{equation}
The relative sign between the two propagators in Eq.~\ref{eq:cq_dd} follows from the orthogonal mixing of the singlet and doublet pseudoscalar states. The overall interaction is proportional to $y_\chi\sin\theta\cos\theta$, reflecting the fact that the mediator must contain both a singlet component to couple to DM, and a doublet component to couple to SM quarks.

At the nucleon level, this interaction maps onto the momentum-dependent non-relativistic effective field theory (NREFT) operator $\mathcal{O}_6^{\rm NR}$. In the conventions adopted here, the effective interaction can be written as
\begin{equation}
\mathcal{L}_{\rm eff}
=
4c_N(Q^2)\mathcal{O}_6^{\rm NR}
=
4c_N(Q^2)
\left(\mathbf{S}_\chi\cdot\mathbf{Q}\right)
\left(\mathbf{S}_N\cdot\mathbf{Q}\right),
\label{eq:O6_NREFT}
\end{equation}
where $\mathbf{S}_\chi$ and $\mathbf{S}_N$ denote the DM and nucleon spins, respectively. The corresponding nucleon Wilson coefficient can be written as \cite{Dolan:2014ska}
\begin{align}
c_N(Q^2)
&=
\sum_{q=u,d,s}
\frac{m_N}{m_q}
\left[
c_q(Q^2)
-
\sum_{q'=u,d,s,c,b,t}
c_{q'}(Q^2)
\frac{\overline{m}}{m_{q'}}
\right]
\Delta_q^N,
\nonumber\\
\overline{m}
&=
\left(
\frac{1}{m_u}
+
\frac{1}{m_d}
+
\frac{1}{m_s}
\right)^{-1}.
\label{eq:cN_matching}
\end{align}
Here $\Delta_q^N$ denotes the contribution of the quark $q$ to the spin of the nucleon $N$.

The corresponding differential DM--nucleus scattering cross section is given by \cite{Arina:2014yna,Dolan:2014ska}
\begin{equation}
\frac{d\sigma_T}{dE_R}
=
\frac{m_T}{32\pi v_E^2}
\frac{Q^4}{m_N^2m_\chi^2}
\sum_{N,N'=p,n}
g_N g_{N'}
F_{\Sigma^{\prime\prime}}^{NN'}(Q^2),
\label{eq:diff_cross_section}
\end{equation}
where $v_E$ is the DM speed in the Earth frame, $g_N$ denotes the corresponding effective coupling to a proton or neutron, and $F_{\Sigma^{\prime\prime}}^{NN'}$ are the nuclear response functions associated with the longitudinal spin response. For xenon, the dominant contributions arise from the naturally abundant isotopes $^{129}{\rm Xe}$ and $^{131}{\rm Xe}$, which carry non-zero nuclear spin. We use the response functions defined in Ref.~\cite{Fitzpatrick:2012ix}.

Equation~\ref{eq:diff_cross_section} makes explicit the characteristic $Q^4$ dependence generated by $\mathcal{O}_6^{\rm NR}$. This strong momentum dependence is central to the present analysis. In contrast to conventional SI scattering, whose recoil spectrum is strongly concentrated toward low energies, the relative weight of the pseudoscalar-mediated signal increases toward larger momentum transfer. The model can therefore produce a comparatively hard recoil spectrum and accommodate an event at $E_R\simeq248~{\rm keV}$ without necessarily producing a large excess at lower recoil energies.

For elastic scattering, the differential recoil rate can be written as
\begin{equation}
\frac{dR}{dE_R}
=
\frac{N_T\rho_\chi}{m_Tm_\chi}
\int_{v_{\min}}^{v_{\max}}
d^3v\,
f_E(\mathbf{v},t)
\frac{d\sigma_T}{dE_R},
\label{eq:recoil_rate}
\end{equation}
where $N_T$ is the number of target nuclei, $\rho_\chi$ is the local DM density, and $f_E(\mathbf{v},t)$ is the DM velocity distribution in the Earth frame. The minimum incoming DM speed required to produce a recoil energy $E_R$ in an elastic collision is
\begin{equation}
v_{\min}(E_R)
=
\sqrt{
\frac{m_T E_R}{2\mu_{\chi T}^2}
},
\qquad
\mu_{\chi T}
=
\frac{m_\chi m_T}{m_\chi+m_T},
\label{eq:vmin_elastic}
\end{equation}
where $\mu_{\chi T}$ is the DM--nucleus reduced mass.

For the astrophysical inputs, we adopt the fiducial parameters of the Standard Halo Model (SHM) \cite{Drukier:1986tm}. The expected number of signal events is obtained by integrating the predicted recoil spectrum over the LZ analysis window,
\begin{equation}
N_s
=
\mathcal{E}
\int_{E_{R,\min}}^{E_{R,\max}}
dE_R'\,
\epsilon(E_R')
\frac{dR}{dE_R}(E_R'),
\label{eq:lz_counts}
\end{equation}
where $\mathcal{E}$ is the total exposure and $\epsilon(E_R)$ denotes the detector efficiency. We use the LZ exposure $\mathcal{E}=2.84~{\rm tonne\,yr}$ and the recoil-energy interval
\begin{equation}
E_{R,\min}=5.4~{\rm keV},
\qquad
E_{R,\max}=269.9~{\rm keV}.
\end{equation}
and the efficiency $\epsilon(E_R)$ parametrized as in \cite{DiMauro2026}.
We use Eq.~\ref{eq:lz_counts} to determine, for each point in the model parameter space, the value of $y_\chi$ required to reproduce the normalization associated with the LZ candidate event.

\begin{figure}
    \centering
    \includegraphics[width=0.60\linewidth]{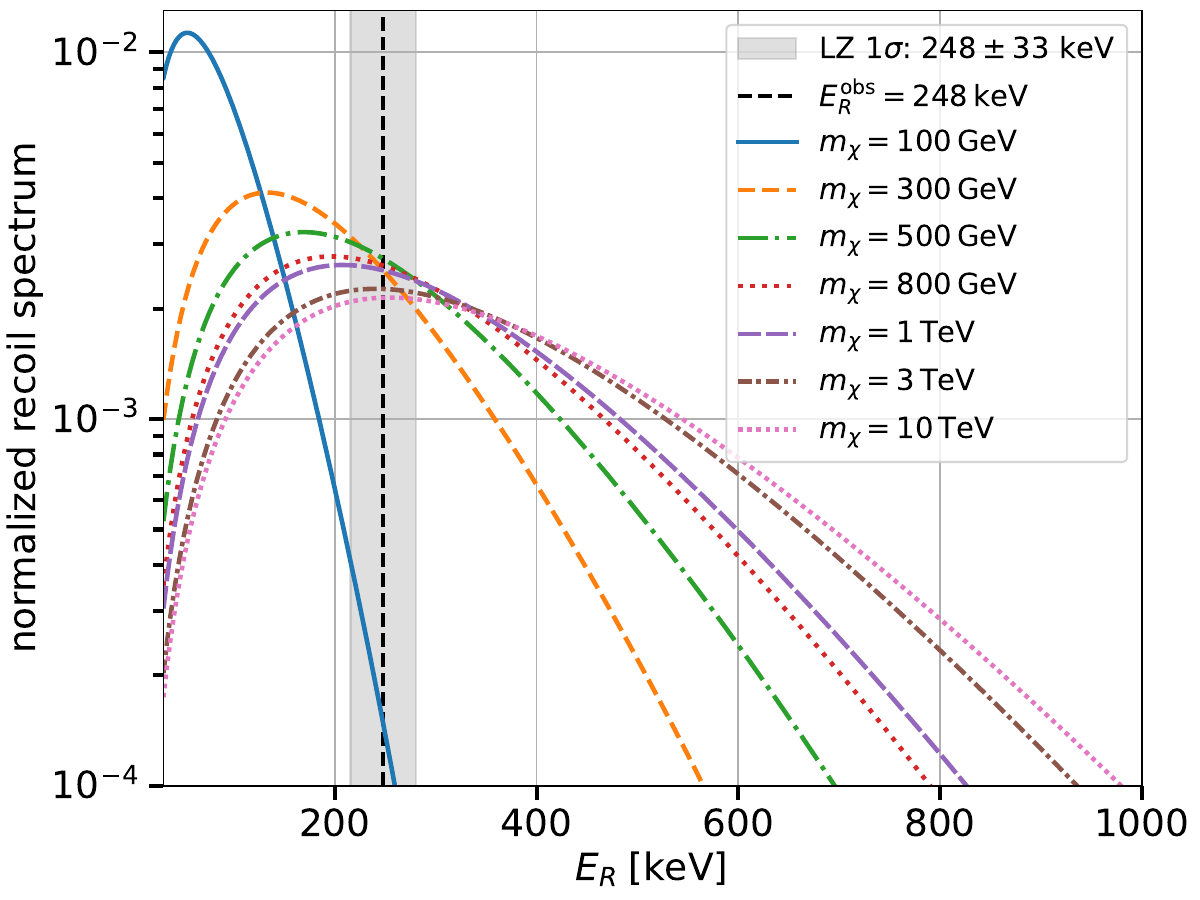}
    \caption{Normalized recoil-energy spectrum for elastic scattering in the benchmark 2HD+a scenario under consideration, shown for representative DM masses $m_\chi=100~{\rm GeV}$, $300~{\rm GeV}$, $500~{\rm GeV}$, $800~{\rm GeV}$, $1~{\rm TeV}$, $3~{\rm TeV}$, and $10~{\rm TeV}$. The vertical dashed line marks the recoil energy of the LZ candidate event, $E_R^{\rm obs}=248~{\rm keV}$, while the gray shaded band indicates the corresponding $1\sigma$ interval, $248\pm33~{\rm keV}$. The spectra become progressively harder as $m_\chi$ increases. In particular, low DM masses predict spectra that fall steeply before reaching the observed recoil energy, whereas heavier DM masses give broader distributions with a substantial high-energy tail and are therefore better suited to account for the LZ event.}
    \label{fig:recoil_spectrum_mass_dependence}
\end{figure}

To better understand the DM-mass dependence of the LZ fit, in Fig.~\ref{fig:recoil_spectrum_mass_dependence} we show the normalized recoil spectrum for a set of representative values of $m_\chi$. The figure makes clear that the shape of the recoil distribution changes significantly with the DM mass. For $m_\chi=100~{\rm GeV}$, the spectrum peaks at relatively low recoil energies and drops rapidly, so that the probability of producing an event around $E_R\simeq248~{\rm keV}$ is strongly suppressed. As the DM mass increases, the recoil spectrum becomes progressively broader and harder, shifting more weight toward large recoil energies. This explains why heavier DM masses are generically more effective at reproducing the high-energy LZ candidate event.
At the same time, Fig.~\ref{fig:recoil_spectrum_mass_dependence} also shows that this effect eventually saturates: once $m_\chi$ reaches the TeV scale, the spectral shape around the LZ event changes more mildly. Therefore, while the high recoil energy favors sufficiently heavy DM, the overall normalization required to reproduce the LZ event still depends on the interplay between $m_\chi$, the mediator masses, the mixing angle, and the coupling $y_\chi$, as discussed in the following figures.

\subsection{Dependence of the LZ normalization on $m_\chi$, $m_a$ and $\tan\beta$}
\label{sec:lz_normalization}

We first investigate how the coupling required by LZ depends on the DM mass and on the parameters controlling the pseudoscalar interaction. For each point considered below, we calculate the full momentum-dependent elastic xenon recoil spectrum, integrate it over the LZ extended recoil-energy window according to Eq.~\ref{eq:lz_counts}, and determine the value of $y_\chi$ for which the predicted number of signal events reproduces the central LZ normalization. Unless otherwise stated, we fix the mass of the heavier pseudoscalar to
\begin{equation}
M_A=800~{\rm GeV},
\end{equation}
and consider representative values of $\tan\beta$.

\begin{figure}
    \centering
    \subfloat{\includegraphics[width=0.33\linewidth]{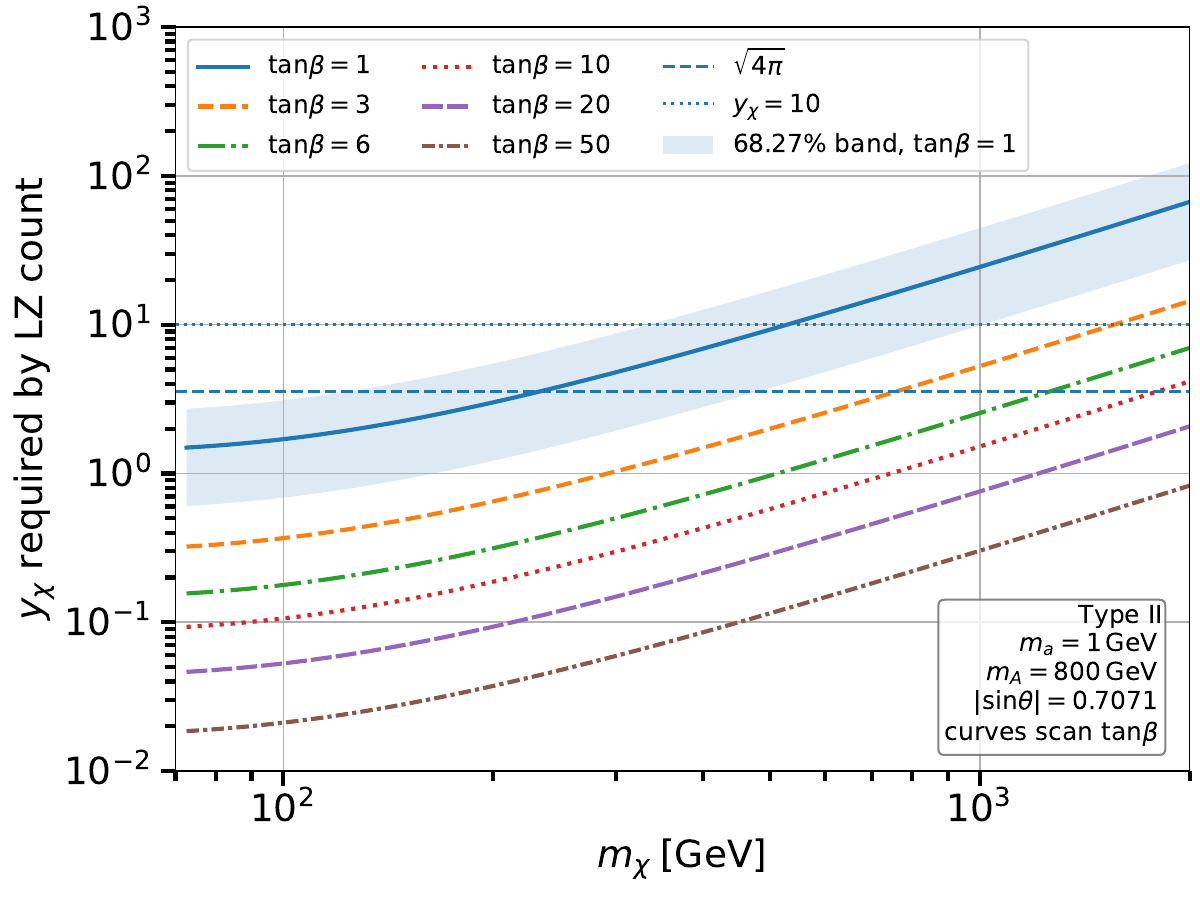}}
    \subfloat{\includegraphics[width=0.33\linewidth]{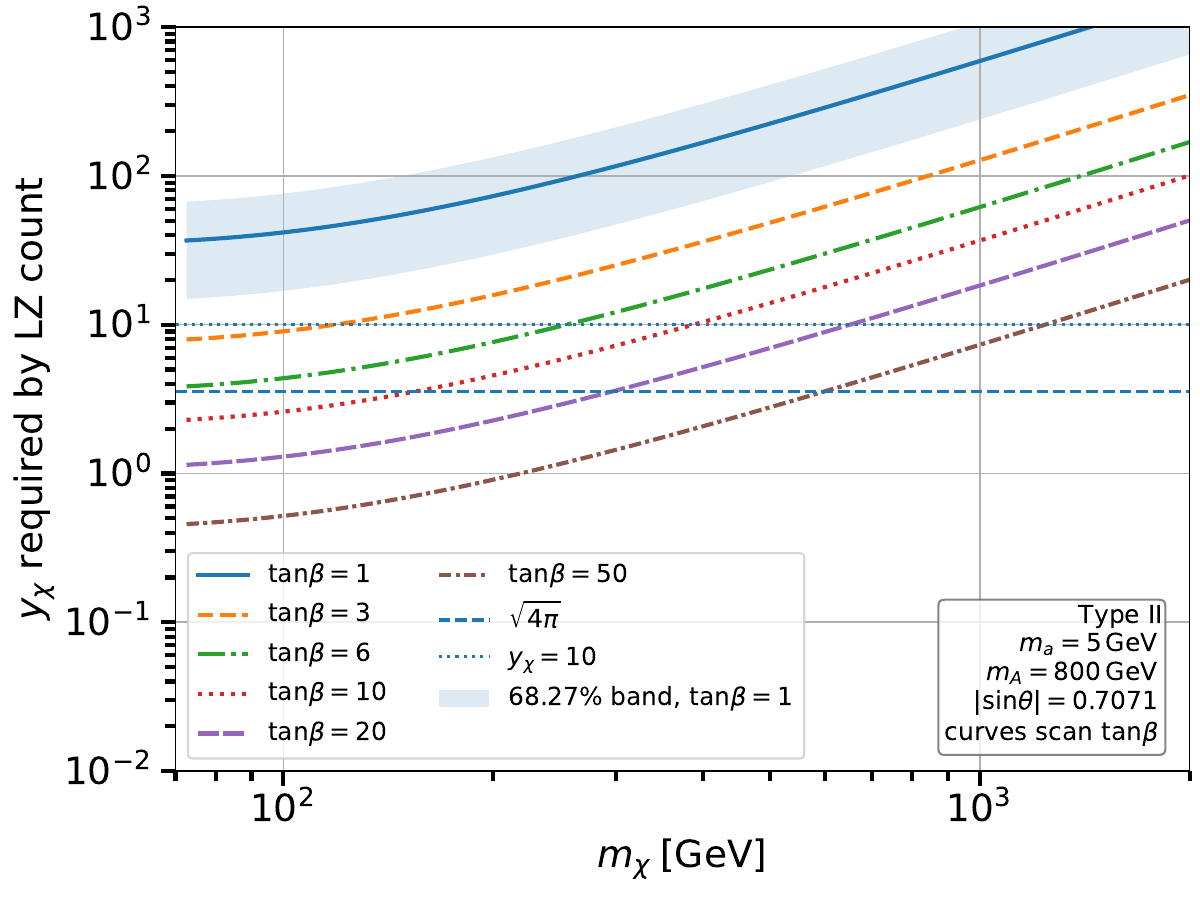}}
    \subfloat{\includegraphics[width=0.33\linewidth]{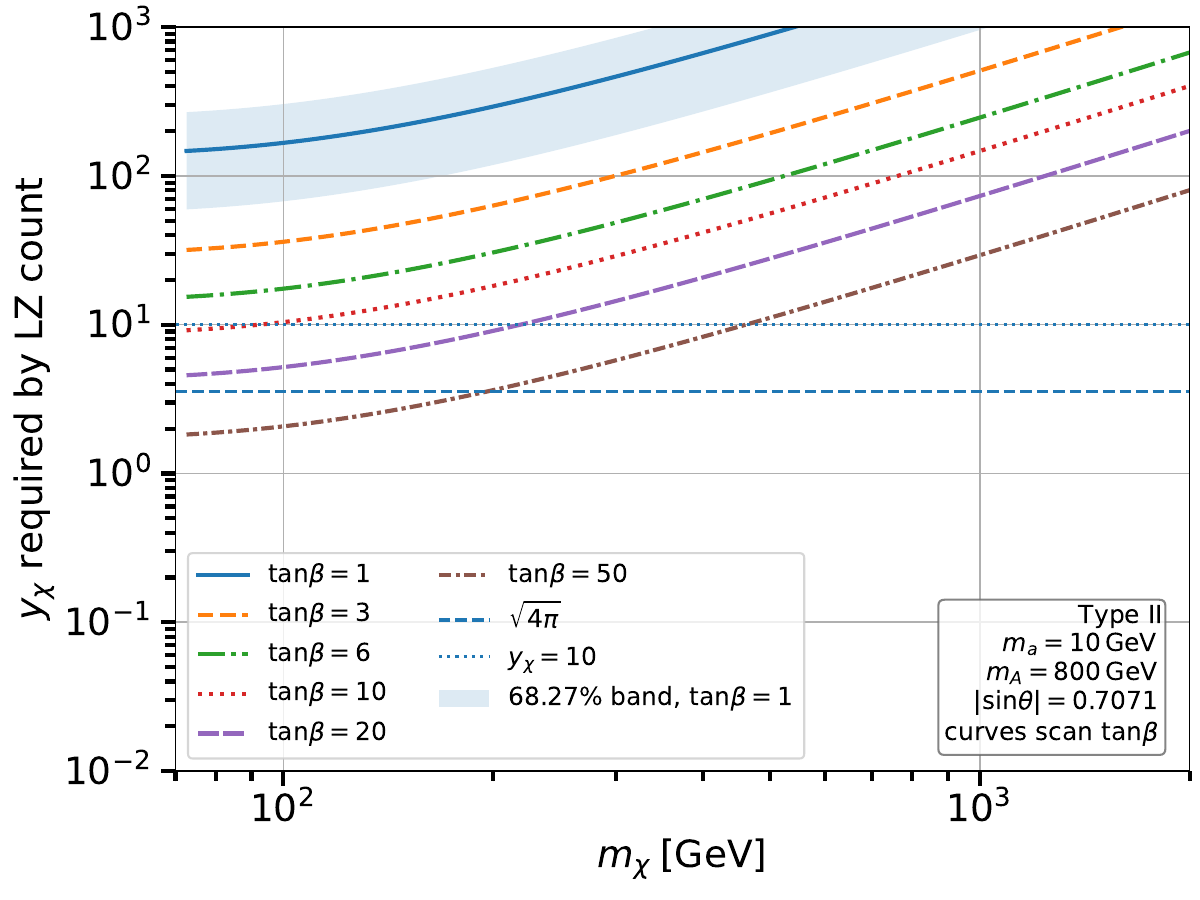}}
    \caption{DM coupling $y_\chi$ required to reproduce the central LZ signal normalization as a function of $m_\chi$ for the Type-II/Type-Y realization of the 2HD+a model. The three panels correspond to $m_a=1~{\rm GeV}$ (left), $m_a=5~{\rm GeV}$ (center), and $m_a=10~{\rm GeV}$ (right), while $M_A=800~{\rm GeV}$ is fixed in all cases. The different curves correspond to $\tan\beta=10$, $15$, $30$, and $50$. The shaded region shows the 68.27\% Poisson interval associated with the one-event normalization for the reference curve. The horizontal dashed and dotted lines indicate $y_\chi=\sqrt{4\pi}$ and $y_\chi=10$, respectively. Decreasing $m_a$ or increasing $\tan\beta$ enhances the scattering rate and therefore reduces the value of $y_\chi$ required by the LZ event.}
    \label{fig:fit_LZ_mchi_ychi}
\end{figure}

As discussed in Sec.~\ref{sec:model}, Type-II and Type-Y have identical quark Yukawa couplings and therefore give the same tree-level DM--nucleus scattering rate. The same is true for Type-I and Type-X. The distinction between the two members of each pair becomes important for observables involving charged leptons and will enter the collider and flavor constraints discussed later.
Figure~\ref{fig:fit_LZ_mchi_ychi} shows the value of $y_\chi$ required to reproduce the LZ event as a function of $m_\chi$ for the Type-II/Type-Y quark-Yukawa structure. The three panels correspond to
\begin{equation}
m_a=10~{\rm GeV},
\qquad
m_a=5~{\rm GeV},
\qquad
m_a=1~{\rm GeV},
\end{equation}
from the upper-left to the lower panel, respectively. In each panel, the curves correspond to $\tan\beta=1$, $3$, $6$, $10$, $20$, and $50$. The vertical axis therefore directly gives, for each combination of $m_\chi$, $m_a$, and $\tan\beta$, the value of the DM coupling required to reproduce the central LZ signal normalization.

The dependence on $\tan\beta$ is particularly transparent in the Type-II/Type-Y realization. The pseudoscalar coupling to down-type quarks scales as
\begin{equation}
g_{Add}\propto\tan\beta,
\label{eq:gadd_tanbeta}
\end{equation}
whereas the coupling to up-type quarks scales as $\cot\beta$. In the parameter region relevant for xenon scattering, increasing $\tan\beta$ therefore enhances the down-type contribution to the nuclear scattering amplitude. Consequently, a smaller value of $y_\chi$ is required to obtain the same event rate. This explains the ordering of the curves in Fig.~\ref{fig:fit_LZ_mchi_ychi}: at fixed $m_\chi$ and $m_a$, $\tan\beta=50$ requires the smallest DM coupling, while $\tan\beta=1$ requires a substantially larger value.

The dependence on the light-pseudoscalar mass is even more pronounced. The tree-level scattering amplitude contains the propagator combination
\begin{equation}
{\cal M}_{\rm DD}
\propto
y_\chi\sin\theta\cos\theta\,
g_{Aqq}
\left[
\frac{1}{Q^2+m_a^2}
-
\frac{1}{Q^2+M_A^2}
\right].
\label{eq:DD_scaling_ma}
\end{equation}

For $M_A\gg m_a$, the contribution of the heavy pseudoscalar is suppressed, and the amplitude is predominantly controlled by the light state $a$. Lowering $m_a$ therefore enhances the light-mediator contribution and substantially reduces the value of $y_\chi$ required to reproduce the event. This trend can be seen directly by comparing the three panels of Fig.~\ref{fig:fit_LZ_mchi_ychi}. For $m_a=10~{\rm GeV}$, rather large values of $y_\chi$ are required over much of the DM mass range. The required coupling decreases significantly for $m_a=5~{\rm GeV}$ and becomes of order unity, or even below unity in part of the parameter space, for $m_a=1~{\rm GeV}$. The LZ normalization therefore already indicates a clear preference for a relatively light pseudoscalar mediator.

For fixed $m_a$ and $\tan\beta$, the coupling required by LZ generally increases with $m_\chi$ over the mass range shown. This behavior results from the interplay between recoil kinematics, the xenon nuclear response, and the DM flux. Although a heavier DM particle can more easily generate a large nuclear recoil, its local number density decreases approximately as $1/m_\chi$. Over the range displayed in Fig.~\ref{fig:fit_LZ_mchi_ychi}, an increasingly large interaction strength is therefore required to maintain an expected event rate of order unity.

The horizontal reference lines at
\begin{equation}
y_\chi=\sqrt{4\pi}
\end{equation}
and
\begin{equation}
y_\chi=10
\end{equation}
provide useful indications of the coupling strength. The first is adopted as a conventional reference for the onset of a strongly coupled regime, rather than as a sharp theoretical boundary, while the second indicates the broader range explored in our numerical scan. Their intersections with the LZ-required curves provide an immediate indication of which combinations of $m_\chi$, $m_a$, and $\tan\beta$ can reproduce the candidate event with a moderately coupled dark sector.

The shaded region in Fig.~\ref{fig:fit_LZ_mchi_ychi} illustrates the sizeable statistical uncertainty associated with interpreting a single observed event. It corresponds to the 68.27\% Poisson interval for the expected signal normalization. Since the tree-level number of events scales as
\begin{equation}
N_s\propto y_\chi^2,
\end{equation}
the uncertainty on a one-event signal translates into a relatively broad interval in the coupling. This uncertainty becomes particularly relevant when the direct-detection requirement is compared with the independent constraints discussed below and, in particular, with the value of $y_\chi$ favored by the thermal relic abundance.

\begin{figure}
    \centering
    \includegraphics[width=0.45\linewidth]{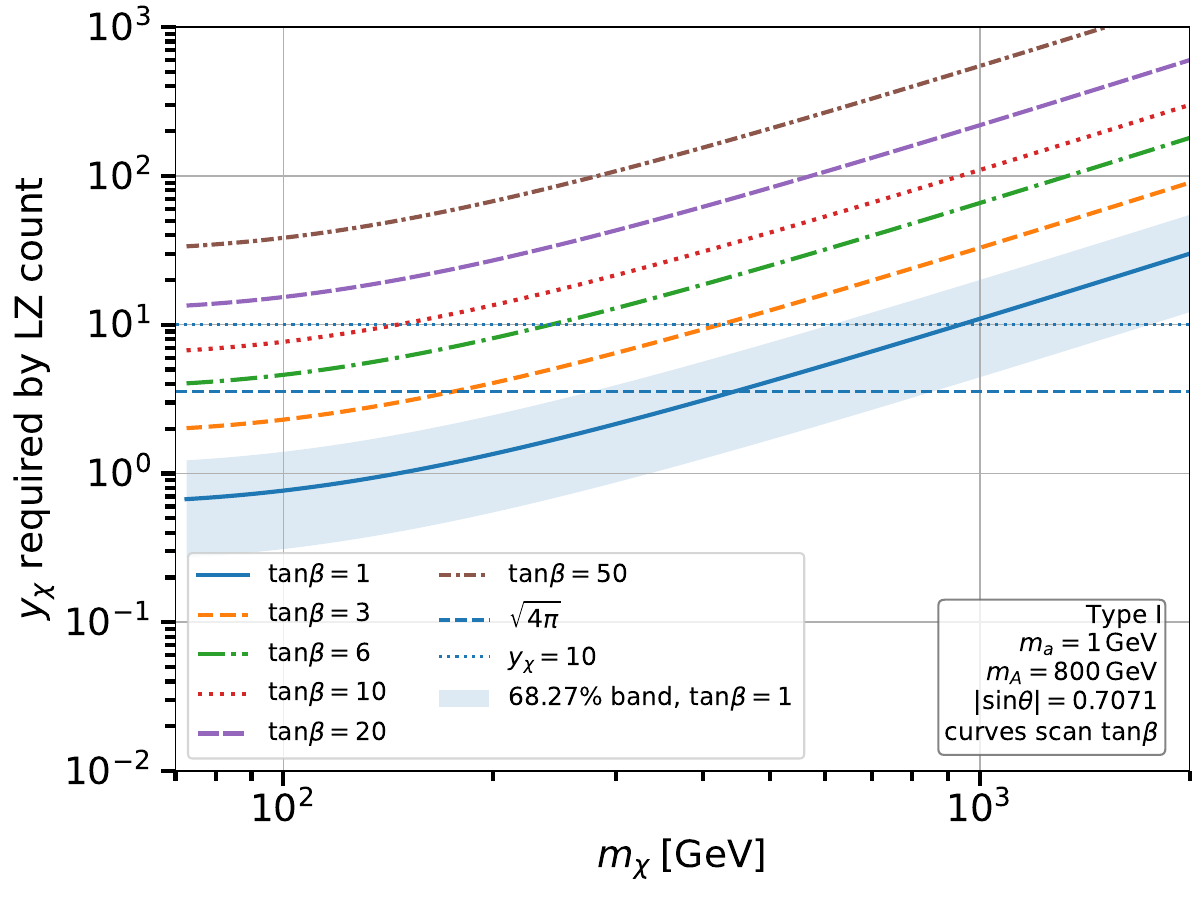}
    \caption{DM coupling $y_\chi$ required to reproduce the central LZ signal normalization as a function of $m_\chi$ for the Type-I/Type-X realization. We fix $m_a=1~{\rm GeV}$ and $M_A=800~{\rm GeV}$ and show $\tan\beta=1$, $3$, $6$, $10$, $20$, $50$. In contrast to the Type-II/Type-Y case, the pseudoscalar couplings to quarks scale as $\cot\beta$, and the value of $y_\chi$ required by LZ therefore increases with $\tan\beta$. The shaded region shows the 68.27\% one-count Poisson interval for the $\tan\beta=1$ benchmark, while the horizontal lines indicate $y_\chi=\sqrt{4\pi}$ and $y_\chi=10$. Only the light-mediator benchmark $m_a=1~{\rm GeV}$ is displayed because larger pseudoscalar masses require substantially larger values of $y_\chi$.}
    \label{fig:fit_LZ_mchi_ychi_bis}
\end{figure}

Figure~\ref{fig:fit_LZ_mchi_ychi_bis} presents the analogous analysis for the Type-I/Type-X realization. The essential difference with respect to Type-II/Type-Y is the dependence of the quark couplings on $\tan\beta$. In Type-I and Type-X, both the up-type and down-type quark couplings scale in magnitude as
\begin{equation}
|g_{Aqq}|\propto\cot\beta.
\label{eq:typeI_cotbeta}
\end{equation}
The tree-level nuclear scattering amplitude therefore decreases as $\tan\beta$ is increased. At fixed $m_\chi$ and $m_a$, reproducing the same LZ event rate consequently requires an increasingly large $y_\chi$ as $\tan\beta$ grows. The ordering of the curves is thus opposite to the Type-II/Type-Y case.

For Type-I/Type-X, we display only the very light-mediator benchmark $m_a=1~{\rm GeV}$. Already for this choice, values of $y_\chi$ above the conventional perturbative reference are required over a substantial fraction of the parameter space, especially for large $m_\chi$ or large $\tan\beta$. Increasing $m_a$ further suppresses the pseudoscalar propagator and pushes the required coupling to even larger values. The Type-I/Type-X interpretation is therefore restricted much more strongly toward a very light pseudoscalar and relatively small $\tan\beta$.

Taken together, Figs.~\ref{fig:fit_LZ_mchi_ychi} and~\ref{fig:fit_LZ_mchi_ychi_bis} illustrate two important features of the elastic 2HD+a interpretation. First, a light pseudoscalar mediator is favored independently of the Yukawa realization. Second, the preferred range of $\tan\beta$ depends qualitatively on the quark-Yukawa structure: large $\tan\beta$ enhances the signal in Type-II/Type-Y, whereas small $\tan\beta$ is favored in Type-I/Type-X.

To illustrate more explicitly the dependence on the parameters controlling the pseudoscalar interaction, we next perform two-dimensional scans of the tree-level elastic-scattering prediction. For every point of these scans, the full momentum-dependent pseudoscalar propagators and xenon nuclear responses are included and the recoil spectrum is integrated over the LZ energy window with the same detector-efficiency prescription adopted above. Unless otherwise stated, we fix
\begin{equation}
m_\chi=120~{\rm GeV},
\qquad
M_A=800~{\rm GeV}.
\end{equation}

\begin{figure}
    \centering
    \subfloat{\includegraphics[width=0.5\linewidth]{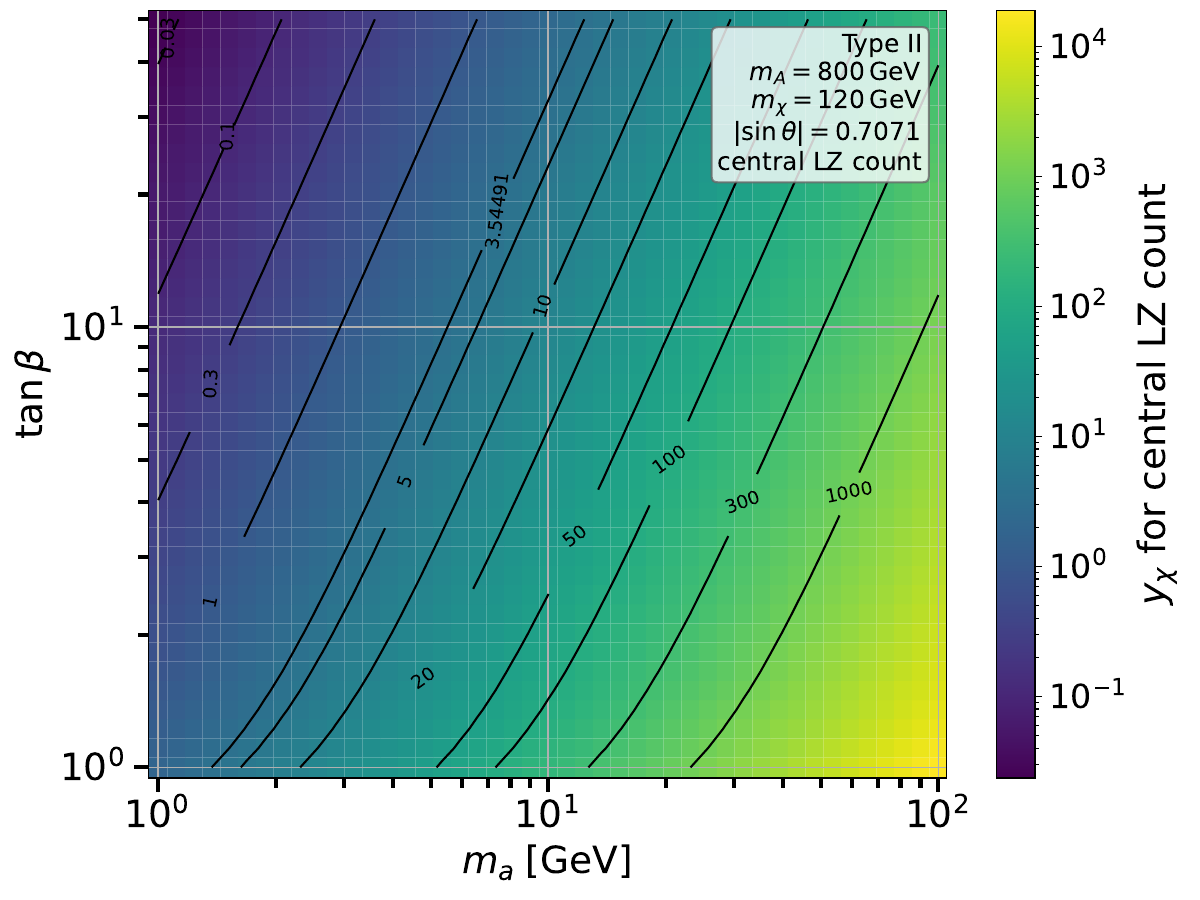}}
    \subfloat{\includegraphics[width=0.5\linewidth]{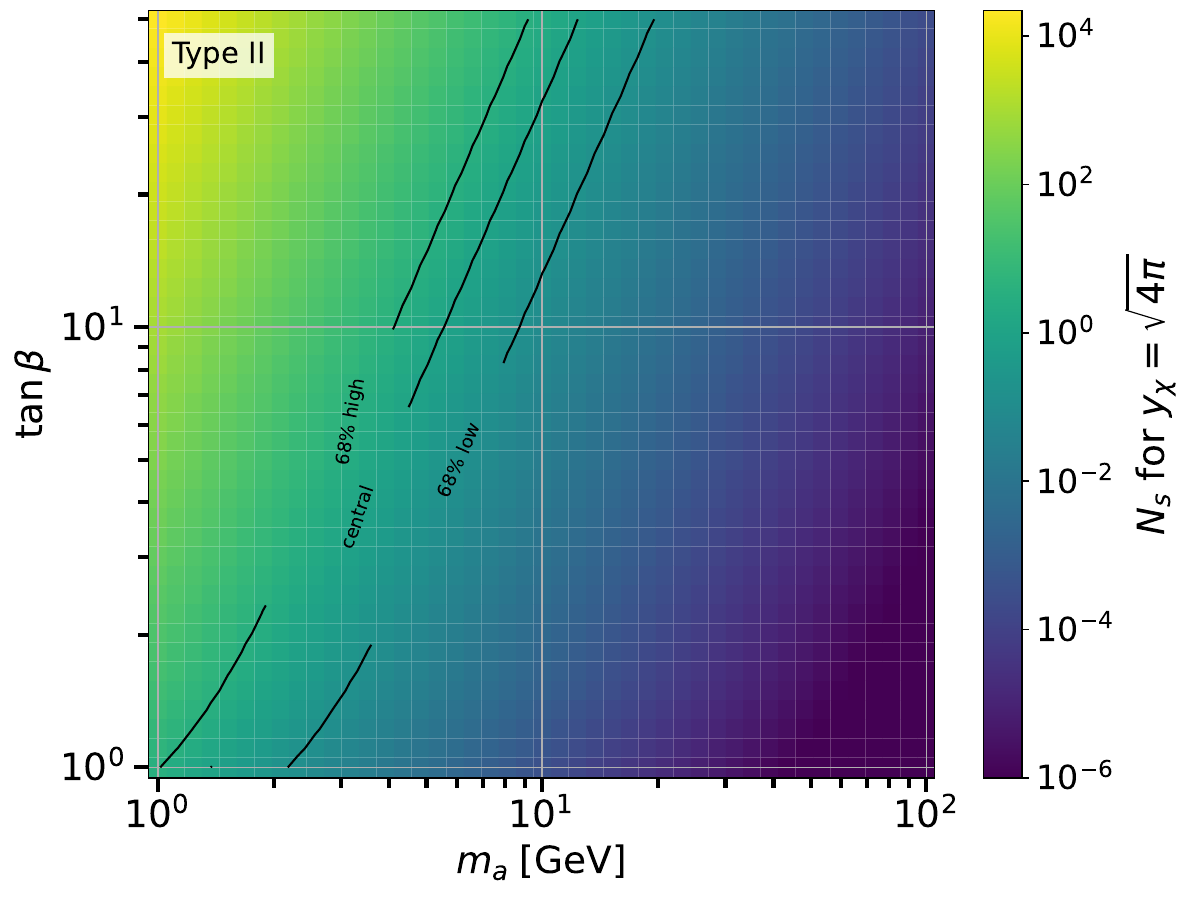}}
    \caption{Dependence of the LZ signal normalization on the light-pseudoscalar mass $m_a$ and on $\tan\beta$ for the Type-II/Type-Y quark-Yukawa structure. We fix $m_\chi=120~{\rm GeV}$, $M_A=800~{\rm GeV}$, and $|\sin\theta|=1/\sqrt{2}$. Left: value of the DM coupling $y_\chi$ required to reproduce the central LZ signal normalization in the $[m_a,\tan\beta]$ plane. The black curves denote contours of constant $y_\chi$. Right: expected number of LZ signal events $N_s$ for the fixed reference coupling $y_\chi=\sqrt{4\pi}$. The black contours indicate the central one-event normalization and the lower and upper boundaries of the 68.27\% one-count Poisson interval. Smaller $m_a$ and larger $\tan\beta$ both enhance the scattering rate in the Type-II/Type-Y realization.}
    \label{fig:fit_LZ_ma_tanbeta}
\end{figure}

Figure~\ref{fig:fit_LZ_ma_tanbeta} focuses on the dependence on the light-pseudoscalar mass $m_a$ and on $\tan\beta$ for the Type-II/Type-Y quark-Yukawa structure. We additionally fix $|\sin\theta|=1/\sqrt{2}$. In the left panel, the color scale gives the value of $y_\chi$ required to reproduce the central LZ signal normalization at each point in the $[m_a,\tan\beta]$ plane, while the black curves correspond to contours of constant $y_\chi$.
The behavior displayed in the left panel follows directly from the structure of the scattering amplitude. For $M_A\gg m_a$, it is approximately controlled by
\begin{equation}
{\cal M}_{\rm DD}
\propto
y_\chi\sin\theta\cos\theta\,
g_{Aqq}
\left(
\frac{1}{Q^2+m_a^2}
-
\frac{1}{Q^2+M_A^2}
\right).
\end{equation}
Decreasing $m_a$ enhances the light-pseudoscalar propagator and therefore reduces the coupling required by LZ. At the same time, in the Type-II/Type-Y realization the enhancement of the down-type quark coupling at large $\tan\beta$ increases the scattering amplitude. The two effects act in the same direction, explaining why the smallest required values of $y_\chi$ are found for small $m_a$ and large $\tan\beta$. Conversely, the region with a heavier pseudoscalar and small $\tan\beta$ requires increasingly large DM couplings.
The right panel of Fig.~\ref{fig:fit_LZ_ma_tanbeta} shows the same parameter plane from a complementary perspective. Rather than solving for $y_\chi$ at each point, we fix the coupling to the reference value $y_\chi=\sqrt{4\pi}$
which corresponds to the conventional perturbativity criterion $y_\chi^2/(4\pi)\lesssim$. We then compute the corresponding number of expected LZ signal events, $N_s$, allowing us to identify directly which regions of parameter space can reproduce the observed event without requiring the dark-sector coupling to enter a strongly coupled regime. The color scale therefore directly indicates whether a coupling of this size underproduces or overproduces the observed event. The black contours identify the central one-event normalization and the lower and upper boundaries of the 68.27\% Poisson interval. The two panels thus encode the same direct-detection information in complementary ways: the left panel determines the coupling required by the LZ event, while the right panel identifies the region in which a fixed reference coupling yields a signal normalization compatible with the observed event.

\begin{figure}
    \centering
    \subfloat{\includegraphics[width=0.35\linewidth]{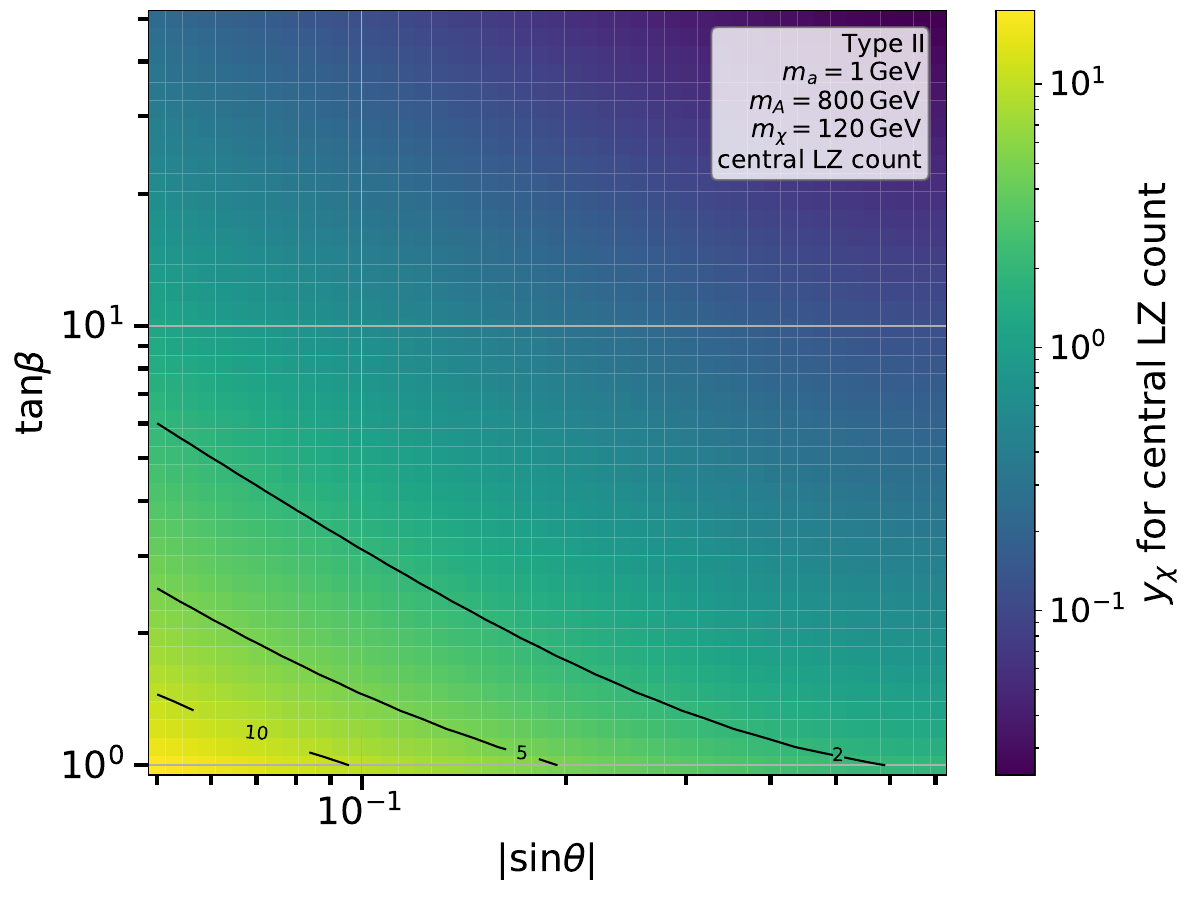}}
    \subfloat{\includegraphics[width=0.35\linewidth]{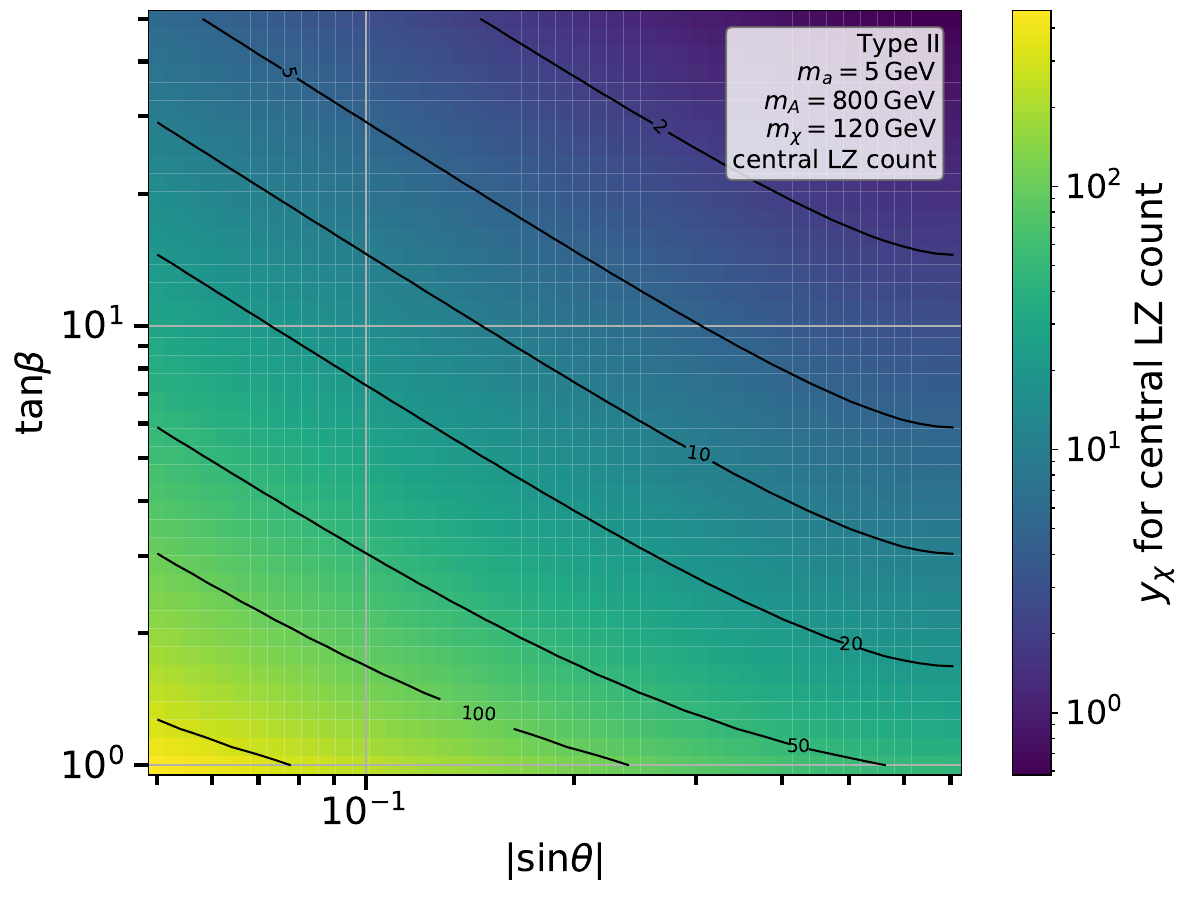}}
    \subfloat{\includegraphics[width=0.35\linewidth]{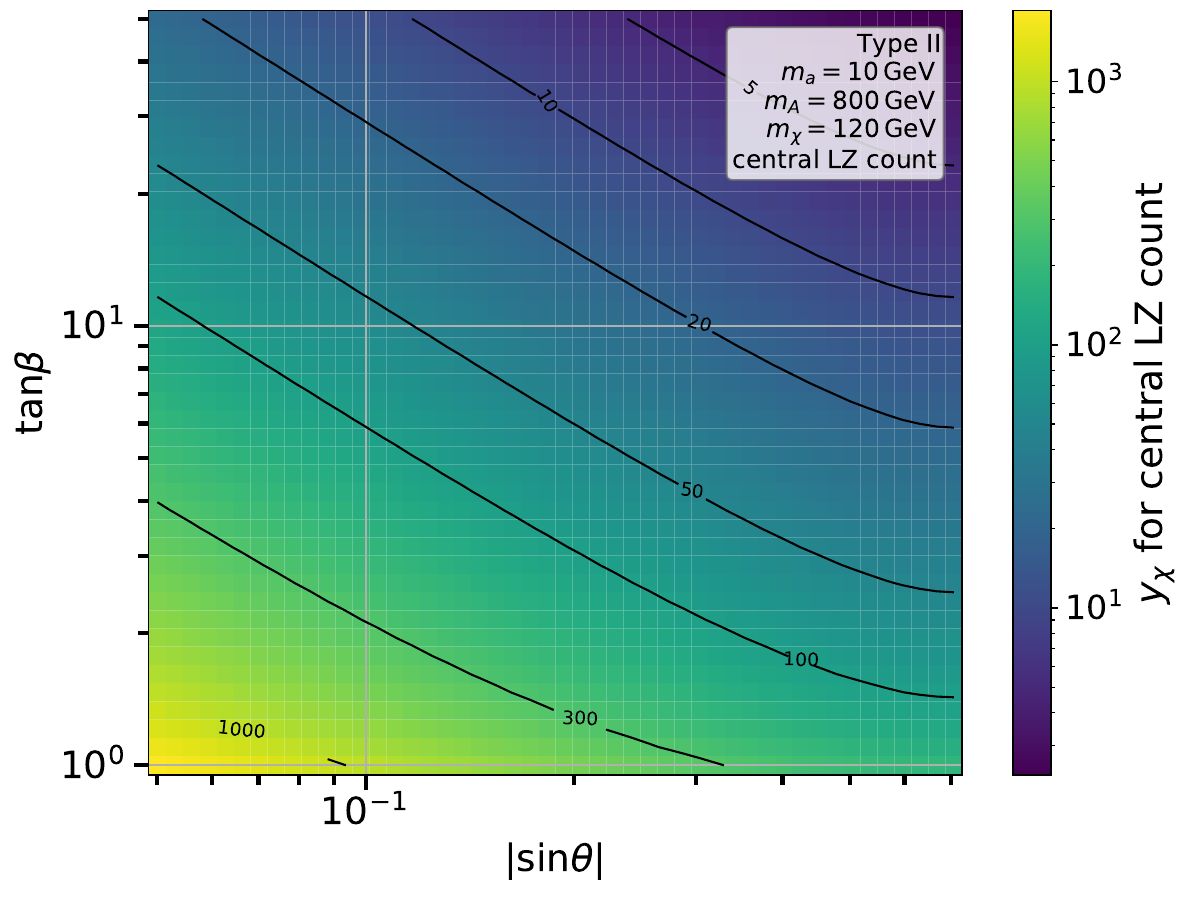}}
    \caption{Dependence of the LZ-required DM coupling $y_\chi$ on the pseudoscalar mixing angle and on $\tan\beta$ for the Type-II/Type-Y quark-Yukawa structure. We fix $m_\chi=120~{\rm GeV}$ and $M_A=800~{\rm GeV}$, while the three panels correspond, from left to right, to $m_a=1$, $5$, and $10~{\rm GeV}$. The color scale gives the value of $y_\chi$ required to reproduce the central LZ signal normalization, and the black curves denote contours of constant coupling. Increasing the singlet--doublet mixing strengthens the interaction connecting the dark and visible sectors, while increasing $\tan\beta$ enhances the down-type quark coupling in Type-II/Type-Y. Increasing $m_a$, on the other hand, suppresses the scattering amplitude and shifts the preferred region toward larger values of $y_\chi$.}
    \label{fig:plot2HDMaTypII_stheta_tbeta}
\end{figure}

The role of the singlet--doublet pseudoscalar mixing is shown in Fig.~\ref{fig:plot2HDMaTypII_stheta_tbeta}. Here we scan the $[|\sin\theta|,\tan\beta]$ plane while keeping $m_\chi=120~{\rm GeV}$ and $M_A=800~{\rm GeV}$ fixed. The three panels correspond, from left to right, to
\begin{equation}
m_a=1~{\rm GeV},
\qquad
m_a=5~{\rm GeV},
\qquad
m_a=10~{\rm GeV}.
\end{equation}
In each panel, the color scale gives the value of $y_\chi$ required to reproduce the central LZ signal normalization, while the black curves denote contours of constant $y_\chi$.

The dependence on the mixing angle follows directly from the structure of the model. The light pseudoscalar must contain a singlet component in order to couple to the DM fermion and a doublet component in order to couple to SM quarks. Consequently, the scattering amplitude is proportional to
\begin{equation}
\sin\theta\cos\theta.
\end{equation}
For small $|\sin\theta|$, the light pseudoscalar is predominantly singlet-like and its coupling to SM quarks is suppressed. A correspondingly larger value of $y_\chi$ is therefore required to reproduce the LZ event. Increasing $|\sin\theta|$ strengthens the communication between the dark and visible sectors and reduces the required coupling. The product $\sin\theta\cos\theta$ reaches its maximum for $|\sin\theta|=\frac{1}{\sqrt{2}}$
corresponding to maximal singlet--doublet mixing. This behavior is clearly visible in all three panels of Fig.~\ref{fig:plot2HDMaTypII_stheta_tbeta}.
The comparison among the three panels also emphasizes the strong impact of the light-pseudoscalar mass. For $m_a=1~{\rm GeV}$, values of $y_\chi$ of order unity, or moderately larger, can reproduce the LZ event over an appreciable region of the $[|\sin\theta|,\tan\beta]$ plane. For $m_a=5~{\rm GeV}$, the required coupling is already substantially larger, while for $m_a=10~{\rm GeV}$ large values of $y_\chi$ are required over most of the displayed parameter space. The mediator mass is therefore one of the key parameters controlling whether the LZ normalization can be obtained within a moderately coupled regime.

Taken together, Figs.~\ref{fig:fit_LZ_mchi_ychi}--\ref{fig:plot2HDMaTypII_stheta_tbeta} identify the basic conditions under which the tree-level pseudoscalar interaction can reproduce the LZ candidate event. The direct-detection normalization favors a relatively light pseudoscalar mediator, sizable singlet--doublet mixing, and a value of $\tan\beta$ that enhances the relevant quark coupling. This implies large $\tan\beta$ for Type-II/Type-Y and small $\tan\beta$ for Type-I/Type-X. These conclusions follow from the tree-level LZ normalization alone. We next investigate whether the same parameter regions remain viable once the loop-induced SI interaction and the complementary collider, flavor, and cosmological constraints are imposed.

\subsection{Loop-level direct detection}
\label{sec:dd_loop}

The pseudoscalar interaction discussed above generates the momentum-dependent operator $\mathcal{O}_6^{\rm NR}$ at tree level. A complete assessment of the direct-detection phenomenology must, however, also include SI interactions generated radiatively. These contributions are particularly important because, unlike the tree-level pseudoscalar interaction, they are not suppressed by four powers of the momentum transfer and benefit from the coherent nuclear enhancement characteristic of conventional SI scattering. Consequently, even a loop-suppressed SI amplitude can be strongly constrained by existing low-energy direct-detection searches.

A viable interpretation of the LZ high-energy event therefore requires the loop-induced SI contribution to remain below the current experimental limits throughout the parameter region selected by the tree-level signal. The full loop-induced SI cross section on protons can be written as \cite{Abe:2018emu,Ertas:2019dew,Arcadi:2025sxc,Bell:2026lnk}
\begin{align}
\label{eq:2HDMa_full_loop}
\sigma_{\chi p}^{\rm SI}
&=
\frac{\mu_{\chi p}^2}{\pi}
\frac{m_p^2}{v^2}
\Bigg|
\sum_q
f_q
\sum_{\phi=h,H}
\frac{g_{\phi qq}m_q}{vM_\phi^2}
C_q^{\rm triangle}
+
\sum_{q=u,d,s}
f_q C_{1,q}^{\rm box}
\nonumber\\
&\quad
+
\sum_{q=u,d,s,c,b}
\frac{3}{4}
\left[
q(2)+\bar{q}(2)
\right]
\left[
C_{1,q}^{\rm box}
+
m_\chi C_{2,q}^{\rm box}
\right]
+
\frac{2}{27}
f_{TG}C_G^{\rm box}
\Bigg|^2.
\end{align}
Here $f_q$ and $f_{TG}$ denote the relevant nucleon matrix elements, $q(2)$ and $\bar{q}(2)$ are the second moments of the quark and antiquark parton distribution functions, and $\mu_{\chi p}$ is the DM--proton reduced mass.

The Wilson coefficients $C_q^{\rm triangle}$, $C_{1,q}^{\rm box}$, $C_{2,q}^{\rm box}$, and $C_G^{\rm box}$ encode the contributions from the triangle, box, and gluonic topologies. Their explicit expressions are rather lengthy and are not reproduced here. We use the results collected in Ref.~\cite{Arcadi:2025sxc}, based on the original calculations of Refs.~\cite{Abe:2018emu,Ertas:2019dew}. A revised computation of the SI scattering amplitude has recently been presented in Ref.~\cite{Bell:2026lnk}.

In the numerical analysis presented in the following sections, we require the loop-induced cross section $\sigma_{\chi p}^{\rm SI}$ given in Eq.~\ref{eq:2HDMa_full_loop} to remain below the current leading LZ constraint \cite{LZ2025SI}. This requirement provides an important complementary test of the parameter space capable of producing the high-energy recoil through the tree-level momentum-dependent interaction.

\section{Collider and flavor constraints on the model}
\label{sec:collider}

As discussed in Sec.~\ref{sec:dd}, reproducing the LZ candidate event through the tree-level momentum-dependent interaction favors a relatively light pseudoscalar state $a$. Such a light mediator is, however, subject to several complementary constraints from Higgs measurements, direct collider searches, and flavor observables. In the mass range relevant for our analysis, two collider constraints are particularly important: exotic decays of the SM-like Higgs boson into pseudoscalar pairs and direct searches for light dimuon resonances. We discuss these constraints in turn and subsequently comment on the most relevant flavor bounds.

For
\begin{equation}
m_a<\frac{M_h}{2},
\end{equation}
the decay of the SM-like Higgs boson into a pair of pseudoscalars is kinematically allowed. Its partial width is
\begin{equation}
\Gamma(h\rightarrow aa)
=
\frac{|\lambda_{haa}|^2}{32\pi M_h}
\sqrt{1-\frac{4m_a^2}{M_h^2}},
\label{eq:h_to_aa_width}
\end{equation}
where, in the alignment limit adopted throughout this work, the trilinear coupling is given by
\begin{equation}
\label{eq:tril}
\lambda_{haa}
=
\frac{1}{v}
\left[
\left(
M_h^2+2M_H^2-2m_a^2-2\lambda_3v^2
\right)
\sin^2\theta
-
2
\left(
\lambda_{1P}\cos^2\beta
+
\lambda_{2P}\sin^2\beta
\right)
v^2\cos^2\theta
\right].
\end{equation}
A broad set of searches for exotic Higgs decays into light states has so far found no significant evidence for this topology \cite{ATLAS:2023tkt,CMS:2022qva,ATLAS:2025qyn,ATLAS:2024nnm,ATLAS:2021hbr,ATLAS:2020ahi,CMS:2025hjt,CMS:2024uru}. For the light-pseudoscalar masses considered here, the corresponding bounds require a strong suppression of the $haa$ coupling. We implement this requirement by imposing
\begin{equation}
\frac{|\lambda_{haa}|}{M_h}\lesssim10^{-3}.
\label{eq:haa_constraint}
\end{equation}
Since $\lambda_{haa}$ depends on several independent parameters of the scalar potential, such a suppression can be obtained through appropriate relations among $\lambda_3$, $\lambda_{1P}$, $\lambda_{2P}$, the scalar masses, and the mixing angle. This requirement therefore represents an important restriction on the scalar sector, rather than a constraint that can be expressed in terms of a single model parameter.

To assess its impact together with the other theoretical and experimental requirements discussed in Sec.~\ref{sec:model}, we perform a numerical scan over
\begin{align}
m_\chi &\in [70~{\rm GeV},1~{\rm TeV}],
&
y_\chi &\in [0.01,10],
&
\tan\beta &\in [1,50],
\nonumber\\
M_H,M_A,M_{H^\pm} &\in [M_h,1~{\rm TeV}],
&
|\lambda_3|,|\lambda_{2P}| &\leq 4\pi,
\label{eq:collider_scan}
\end{align}
while $\lambda_{1P}$ is chosen so as to suppress the trilinear coupling $\lambda_{haa}$ according to Eq.~\ref{eq:haa_constraint}. For every scan point, we impose the theoretical consistency conditions, the high-precision constraints discussed in Sec.~\ref{sec:model}, the limit from $h\rightarrow aa$, and the current bounds on loop-induced SI DM scattering discussed in Sec.~\ref{sec:dd_loop}.

\begin{figure}
    \centering
    \subfloat{\includegraphics[width=0.33\linewidth]{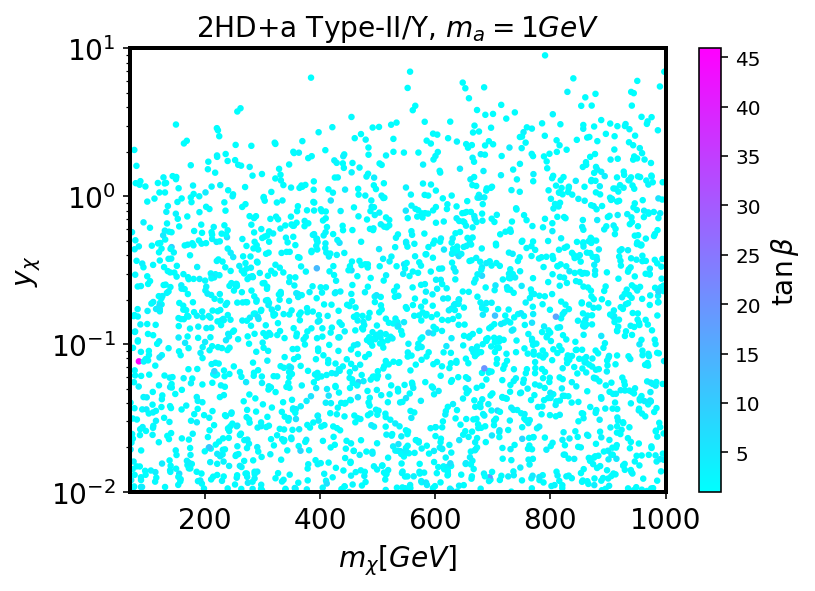}}
    \subfloat{\includegraphics[width=0.33\linewidth]{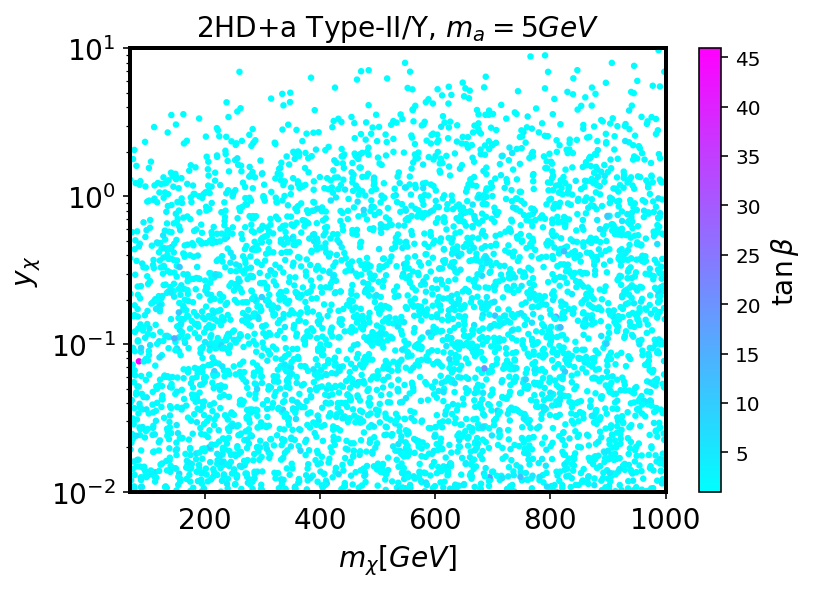}}
    \subfloat{\includegraphics[width=0.33\linewidth]{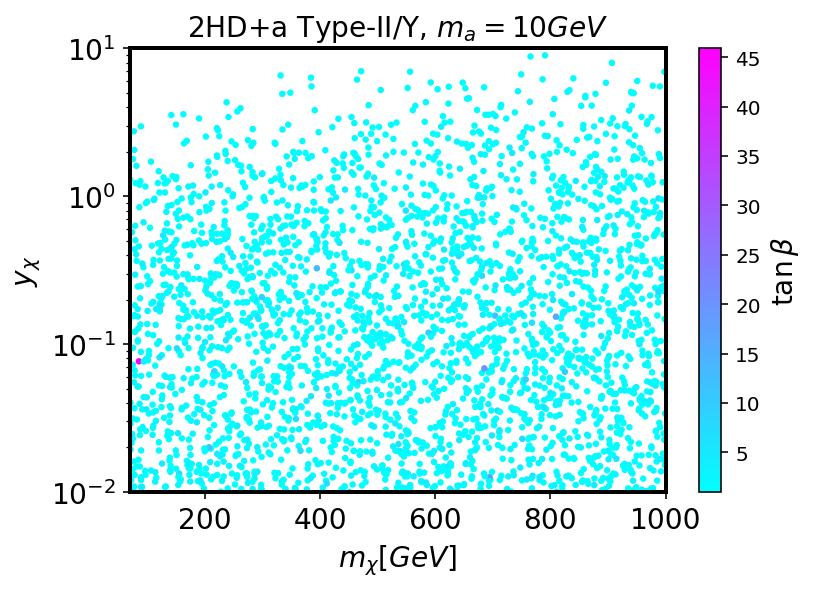}}\\
    \subfloat{\includegraphics[width=0.33\linewidth]{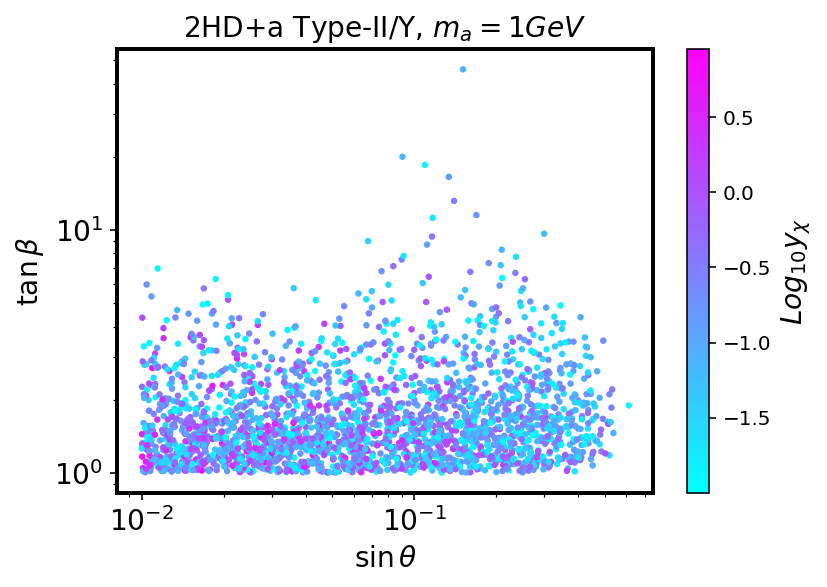}}
    \subfloat{\includegraphics[width=0.33\linewidth]{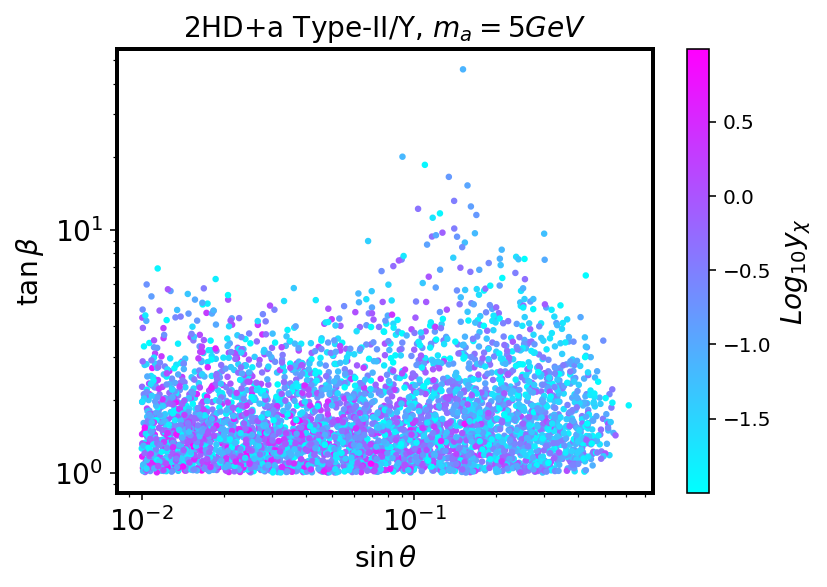}}
    \subfloat{\includegraphics[width=0.33\linewidth]{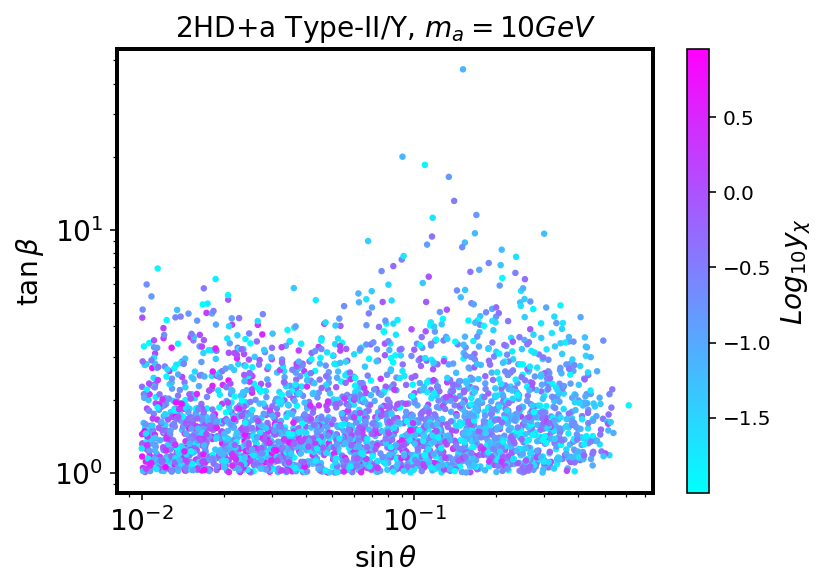}}
    \caption{Parameter points of the 2HD+a model satisfying the theoretical consistency conditions, high-precision constraints, the bound from $h\rightarrow aa$, and the current limits on loop-induced SI DM scattering, for the Type-II/Type-Y quark-Yukawa structure. The three columns correspond to $m_a=1$, $5$, and $10~{\rm GeV}$, from left to right. The upper panels show the accepted points in the $[m_\chi,y_\chi]$ plane, with the color scale indicating $\tan\beta$. The lower panels show the same points in the $[\sin\theta,\tan\beta]$ plane, with the color scale indicating $\log_{10}y_\chi$. The surviving sample is concentrated predominantly at relatively small $\tan\beta$, although isolated configurations remain viable at larger values.}
    \label{fig:TypII_Y_constraintsA}
\end{figure}

Figure~\ref{fig:TypII_Y_constraintsA} shows the surviving parameter points for the Type-II/Type-Y quark-Yukawa structure. The three columns correspond to fixed pseudoscalar masses $m_a=1$, $5$, and $10~{\rm GeV}$. The upper panels display the surviving points in the $[m_\chi,y_\chi]$ plane, with the color scale indicating $\tan\beta$. The lower panels show the same accepted points in the $[\sin\theta,\tan\beta]$ plane, with the color scale indicating $\log_{10}y_\chi$.

The distributions in Fig.~\ref{fig:TypII_Y_constraintsA} show that the complementary constraints significantly reshape the parameter space compared with the LZ requirement alone. While the tree-level LZ normalization in Type-II/Type-Y benefits from large $\tan\beta$, as shown in Sec.~\ref{sec:lz_normalization}, the combination of theoretical, Higgs, precision, and SI direct-detection constraints preferentially retains points at more moderate values of $\tan\beta$. The bulk of the surviving sample lies below approximately $\tan\beta\simeq10$, although a smaller number of configurations survives at larger values. The accepted points also favor moderate pseudoscalar mixing, with most of the viable configurations lying at $|\sin\theta|\lesssim0.5$.

This comparison is important for the interpretation of the LZ event. The parameters that enhance the tree-level momentum-dependent signal are not necessarily those most easily compatible with the complementary constraints. In particular, the lightest mediator benchmark, $m_a=1~{\rm GeV}$, requires substantially smaller values of $y_\chi$ to reproduce the LZ event and therefore offers the largest overlap between the direct-detection requirement and the parameter space surviving the constraints considered in Fig.~\ref{fig:TypII_Y_constraintsA}. For increasing $m_a$, the LZ normalization requires progressively larger values of $y_\chi$, as discussed in Sec.~\ref{sec:lz_normalization}, making the simultaneous satisfaction of all requirements increasingly restrictive.

\begin{figure}
    \centering
    \subfloat{\includegraphics[width=0.5\linewidth]{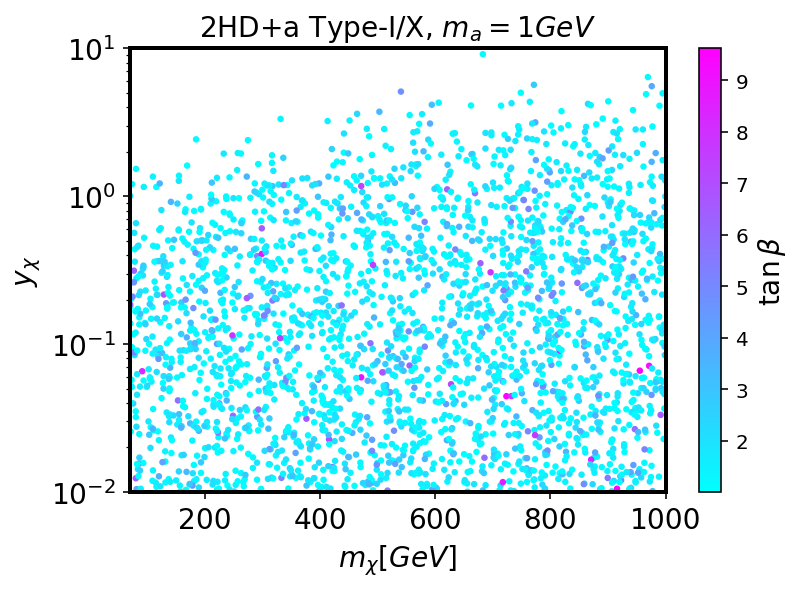}}
    \subfloat{\includegraphics[width=0.5\linewidth]{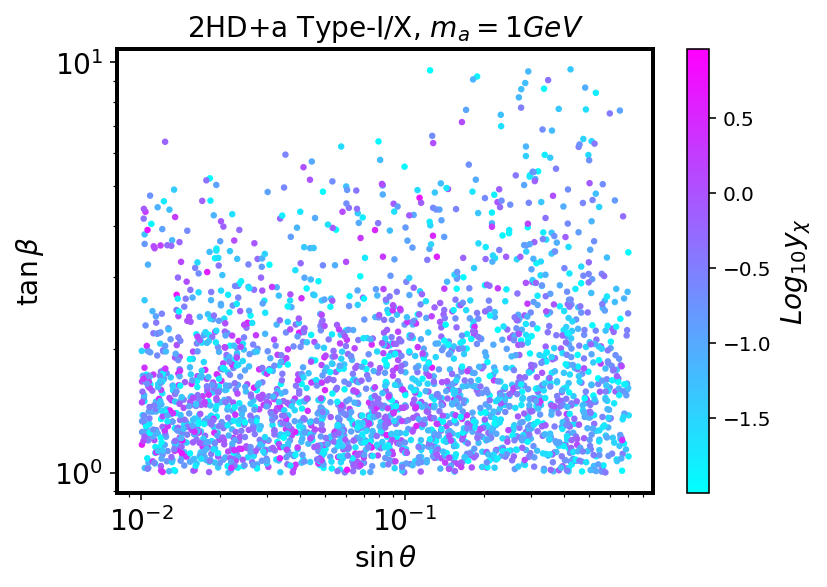}}
    \caption{Same analysis as in Fig.~\ref{fig:TypII_Y_constraintsA}, but for the Type-I/Type-X scenarios. We show only the benchmark $m_a\!=\!1\,{\rm GeV}$ and restrict the scan to $\tan\beta\!<\!10$, since larger $m_a$ or $\tan\beta$ values require increasingly large $y_\chi$ to reproduce the LZ event. The left panel shows the accepted points in the $[m_\chi,y_\chi]$ plane, with the color scale indicating $\tan\beta$, while the right panel shows the same points in the $[\sin\theta,\tan\beta]$ plane, with the colors indicating $\log_{10}y_\chi$.}
    \label{fig:TypI_X_constraintsA}
\end{figure}

Figure~\ref{fig:TypI_X_constraintsA} presents the analogous analysis for the Type-I/Type-X quark-Yukawa structure. As already found from the LZ normalization alone, these realizations are significantly more restrictive because all pseudoscalar couplings to quarks decrease as $\cot\beta$. We therefore restrict the scan to $\tan\beta<10$ and display only the benchmark $m_a=1~{\rm GeV}$, since larger pseudoscalar masses require increasingly large values of $y_\chi$ to reproduce the LZ event.

The comparison between Figs.~\ref{fig:TypII_Y_constraintsA} and~\ref{fig:TypI_X_constraintsA} highlights the different role played by $\tan\beta$ in the two quark-Yukawa structures. In Type-II/Type-Y, increasing $\tan\beta$ enhances the down-type quark coupling and therefore the tree-level LZ scattering rate. In Type-I/Type-X, instead, all quark couplings are suppressed as $\cot\beta$, so the LZ interpretation becomes rapidly more difficult as $\tan\beta$ is increased. Consequently, the Type-I/Type-X explanation is confined to a significantly smaller region characterized by a very light pseudoscalar, relatively small $\tan\beta$, and sufficiently large $y_\chi$.

A second particularly important collider constraint in the pseudoscalar mass window relevant for the LZ interpretation arises from direct searches for light dimuon resonances,
\begin{equation}
pp\rightarrow a\rightarrow\mu^+\mu^-.
\end{equation}
At hadron colliders, the pseudoscalar can be produced through gluon fusion,
\begin{equation}
gg\rightarrow a,
\end{equation}
mediated primarily by heavy-quark loops, or through bottom-quark fusion,
\begin{equation}
b\bar{b}\rightarrow a,
\end{equation}
when the coupling of $a$ to bottom quarks is enhanced. For very light pseudoscalars, charm-quark contributions to gluon fusion can also become relevant. Searches for narrow dimuon resonances have been performed over complementary mass ranges by ATLAS \cite{ATLAS:2026bpi}, CMS \cite{CMS:2023hwl}, and LHCb \cite{LHCb:2020ysn}.

\begin{figure}
    \centering
    \subfloat{\includegraphics[width=0.45\linewidth]{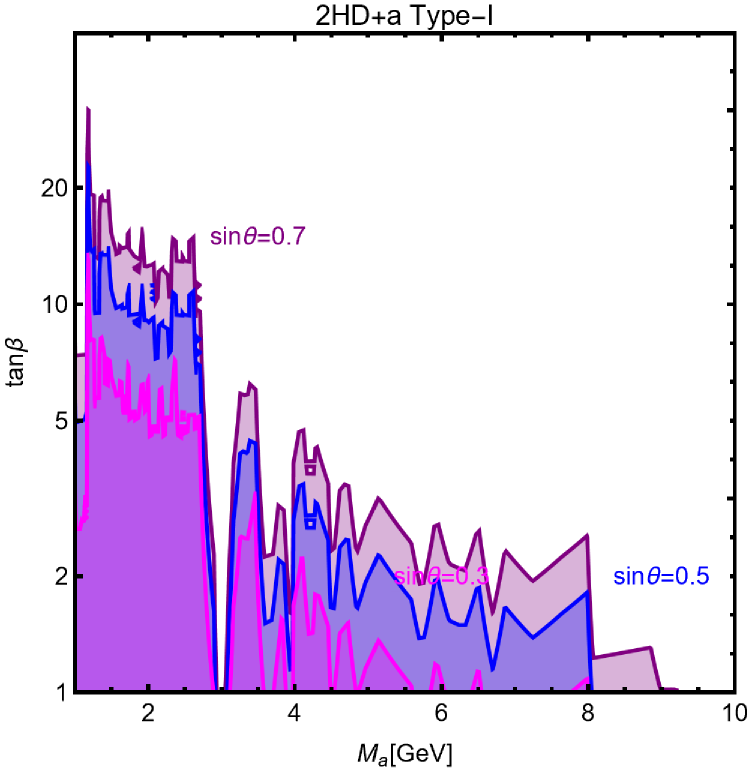}}~~~
    \subfloat{\includegraphics[width=0.45\linewidth]{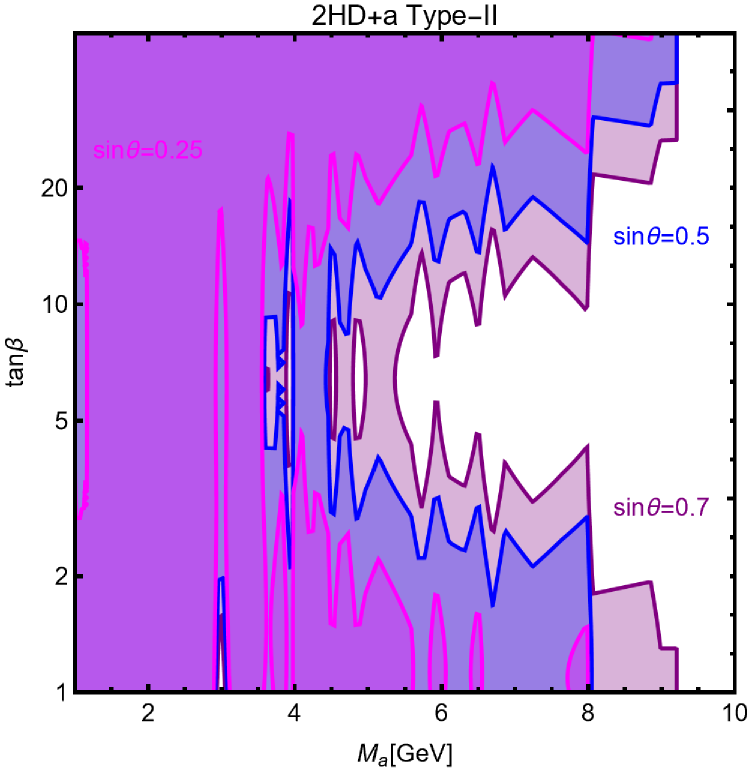}}\\
    \subfloat{\includegraphics[width=0.45\linewidth]{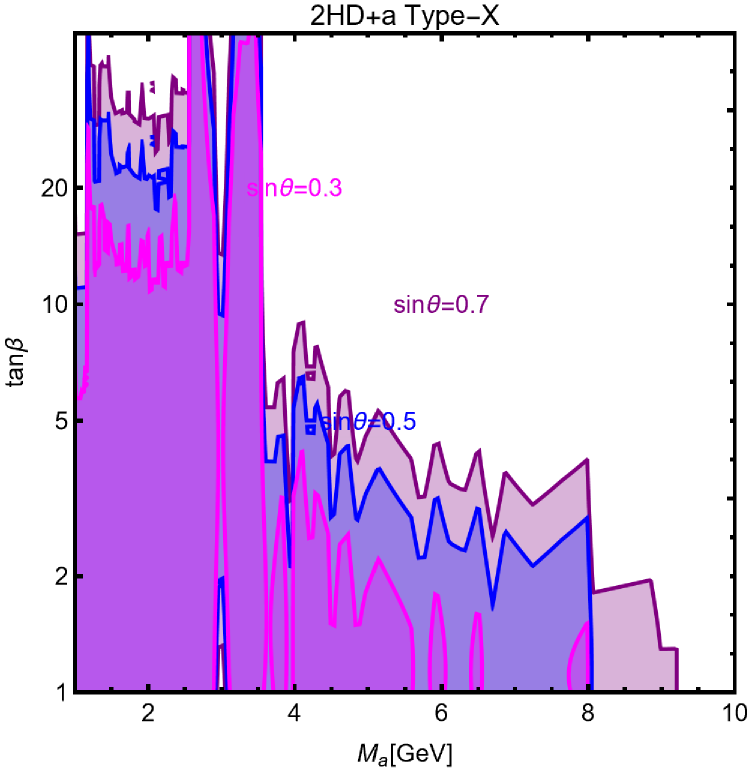}}~~~
    \subfloat{\includegraphics[width=0.45\linewidth]{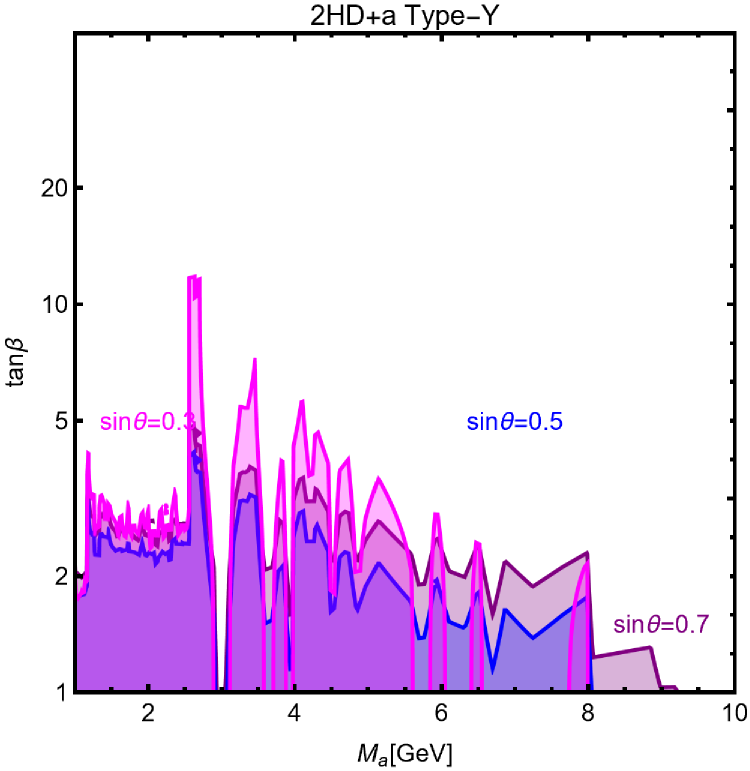}}
    \vspace*{-2mm}
    \caption{Exclusion regions derived from searches for the process $pp\rightarrow a\rightarrow\mu^+\mu^-$. The four panels correspond to the Type-I, Type-II, Type-X, and Type-Y Yukawa realizations of the 2HD+a model. The excluded regions are shown in the $[m_a,\tan\beta]$ plane for representative values of the pseudoscalar mixing angle $\sin\theta$, indicated in each panel. The different behavior among the four realizations reflects the distinct $\tan\beta$ dependence of the pseudoscalar couplings to quarks and charged leptons.}
    \label{fig:pLHC}
\vspace*{-2mm}
\end{figure}

We recast the CMS search for a singlet extension of the 2HDM \cite{CMS:2023hwl} for the four FCNC-preserving Yukawa realizations of the 2HD+a model. The resulting exclusion regions in the $[m_a,\tan\beta]$ plane are shown in Fig.~\ref{fig:pLHC}. The four panels correspond to Type-I, Type-II, Type-X, and Type-Y, while the differently colored regions show the parameter space excluded for representative values of the pseudoscalar mixing angle $\sin\theta$.

The impact of the dimuon searches depends strongly on the Yukawa realization. In Type-II, both the bottom-quark and charged-lepton couplings are enhanced at large $\tan\beta$, making this channel particularly constraining. As shown in the upper-right panel of Fig.~\ref{fig:pLHC}, values of $m_a\lesssim6~{\rm GeV}$ are strongly constrained already for moderate $\tan\beta$, while reaching the large-$\tan\beta$ region favored by the tree-level LZ normalization typically requires $m_a\gtrsim8~{\rm GeV}$. This generates a significant tension with the preference of the LZ signal for a light mediator.

The Type-I and Type-X realizations are also strongly affected in the low-mass region. In particular, the very light benchmark $m_a\simeq1~{\rm GeV}$, which is the region most favorable for reproducing the LZ event with manageable values of $y_\chi$, is significantly constrained by the dimuon searches. This further reduces the already limited parameter space identified in Fig.~\ref{fig:TypI_X_constraintsA}.

The situation is qualitatively different in Type-Y. In this case the down-type quark coupling is enhanced at large $\tan\beta$, which is advantageous for the LZ signal, whereas the charged-lepton coupling scales as $\cot\beta$. The branching ratio into muons is therefore suppressed precisely in the large-$\tan\beta$ region in which the DM--nucleus scattering rate is enhanced. As a consequence, the dimuon constraint is substantially weaker than in Type-II, and Type-Y emerges as the most favorable realization among the four Yukawa structures considered here.

In all four scenarios, Fig.~\ref{fig:pLHC} also exhibits a narrow region around $m_a\simeq10~{\rm GeV}$ in which the sensitivity of the dimuon searches deteriorates. This feature originates from the presence of QCD resonances in this mass range, which makes the extraction of a narrow new-physics resonance more challenging. The resulting reduction in experimental sensitivity leaves a small region that is comparatively weakly constrained by the direct dimuon searches.

Finally, light pseudoscalars can also be constrained through flavor processes in which the state $a$ is produced on shell \cite{Dolan:2014ska}. These constraints again depend strongly on the Yukawa realization. In scenarios with an enhanced coupling of $a$ to down-type quarks, such as Type-II and Type-Y, important bounds arise from rare $B$ decays and radiative bottomonium decays, including
\begin{equation}
B\rightarrow K\mu^+\mu^-,
\qquad
\Upsilon\rightarrow\gamma a,
\end{equation}
see, for example, Ref.~\cite{Arcadi:2021zdk}. In Type-II and Type-X, where the pseudoscalar coupling to charged leptons can be enhanced at large $\tan\beta$, the decay
\begin{equation}
B_s\rightarrow\mu^+\mu^-
\end{equation}
provides an additional strong constraint. In particular, pseudoscalar masses below approximately $m_a\lesssim5~{\rm GeV}$ are strongly disfavored over the regions in which the leptonic coupling is enhanced.

The combination of these constraints illustrates the central tension in the elastic 2HD+a interpretation of the LZ event. The direct-detection signal favors a light pseudoscalar and, in the Type-II/Type-Y quark-Yukawa structure, benefits from an enhanced down-type coupling. The same properties, however, increase the sensitivity of exotic Higgs decays, light-resonance searches, and flavor observables. Among the four Yukawa realizations, Type-Y is particularly interesting because it can enhance the coupling to down-type quarks, and hence the LZ scattering rate, without simultaneously enhancing the coupling to charged leptons. Whether the regions surviving these complementary constraints can also reproduce the observed DM abundance is investigated in the next section.

\section{Including the relic density}
\label{sec:relic}

The 2HD+a model can naturally reproduce the observed DM abundance within the conventional thermal WIMP paradigm, since the DM fermion can efficiently annihilate into several SM and Higgs-sector final states. We determine the parameter space compatible with the measured DM relic abundance,
\begin{equation}
\Omega_{\rm DM}h^2 \simeq 0.12,
\end{equation}
using the publicly available package \texttt{micrOMEGAs} \cite{Planck:2018vyg,Belanger:2006is,Belanger:2008sj,Belanger:2013oya}.

To make the main parametric dependences more transparent, we report the leading contributions to the annihilation cross section relevant for thermal freeze-out. At leading order in the non-relativistic velocity expansion, the most important channels can be written as \cite{Arcadi:2019lka,Abe:2018emu,Arcadi:2022lpp,Arcadi:2024ukq}
\begin{align}
\label{eq:2HDMa_sigmav}
\langle\sigma v\rangle_{\bar{f}f}
&=
\frac{1}{2\pi}
\sum_f n_f^c
\sqrt{1-\frac{m_f^2}{m_\chi^2}}\,
y_\chi^2
\sin^2\theta
\cos^2\theta\,
m_\chi^2
\left|
\frac{1}{4m_\chi^2-M_a^2}
-
\frac{1}{4m_\chi^2-M_A^2}
\right|^2,
\nonumber\\
\langle\sigma v\rangle_{XY}
&=
\frac{y_\chi^2}{16\pi}
\sqrt{1-\frac{(M_X+M_Y)^2}{4m_\chi^2}}
\sqrt{1-\frac{(M_X-M_Y)^2}{4m_\chi^2}}
\nonumber\\
&\qquad\times
\left|
\frac{\lambda_{XaY}\cos\theta}{4m_\chi^2-m_a^2}
+
\frac{\lambda_{XAY}\sin\theta}{4m_\chi^2-M_A^2}
\right|^2,
\qquad
X=h,H,
\qquad
Y=a,A,
\nonumber\\
\langle\sigma v\rangle_{aa}
&=
\frac{v_{\rm rel}^2}{12\pi}
\left(
1-\frac{m_a^2}{m_\chi^2}
\right)^{5/2}
y_\chi^4
\frac{m_\chi^6}
{\left(m_a^2-2m_\chi^2\right)^4}.
\end{align}
Here $n_f^c$ denotes the color multiplicity of the final-state fermion and $v_{\rm rel}$ is the relative velocity of the annihilating DM particles. The first contribution corresponds to annihilation into SM fermion pairs through $s$-channel exchange of the two pseudoscalars, while the second represents annihilation into mixed scalar--pseudoscalar final states whenever these channels are kinematically open. The last contribution corresponds to annihilation into a pair of light pseudoscalars and is $p$-wave suppressed, as indicated by its explicit dependence on $v_{\rm rel}^2$.

The relative importance of these channels depends strongly on the DM and mediator masses. Fermionic final states are particularly relevant below the thresholds for additional Higgs states, while new scalar final states can substantially increase the annihilation rate once they become kinematically accessible. In addition, resonant annihilation can occur when $2m_\chi$ approaches the mass of one of the pseudoscalar mediators. The relic-density requirement therefore probes combinations of the same parameters that determine the LZ scattering rate, but through a different dependence on masses, mixing angles, and couplings.

We perform a parameter scan analogous to that described in Sec.~\ref{sec:collider}, now allowing the light-pseudoscalar mass to vary continuously in the range
\begin{equation}
1~{\rm GeV}\leq m_a\leq10~{\rm GeV}.
\end{equation}
The accepted points satisfy the same theoretical, collider, flavor, and loop-induced SI direct-detection constraints discussed previously, together with the requirement that the thermal relic abundance agrees with the observed DM density within the adopted $3\sigma$ interval.

\begin{figure}
    \centering
    \subfloat{\includegraphics[width=0.45\linewidth]{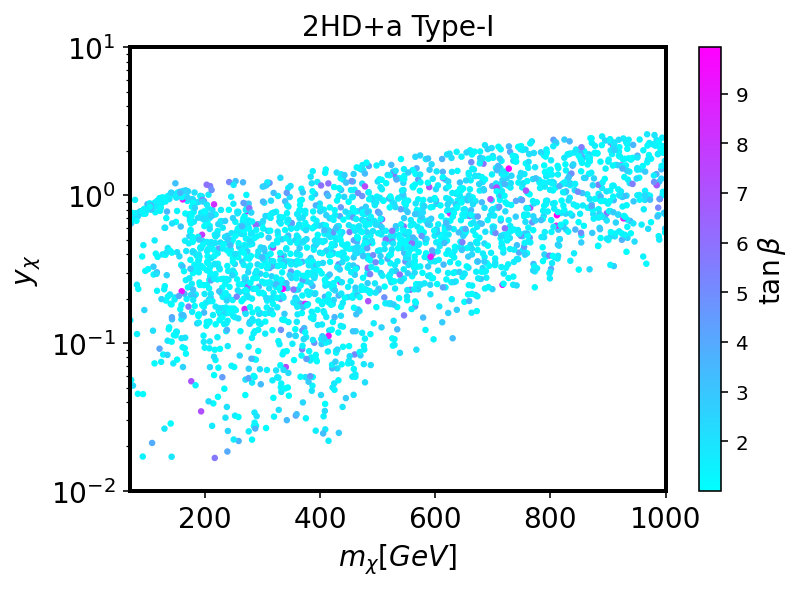}}
    \subfloat{\includegraphics[width=0.45\linewidth]{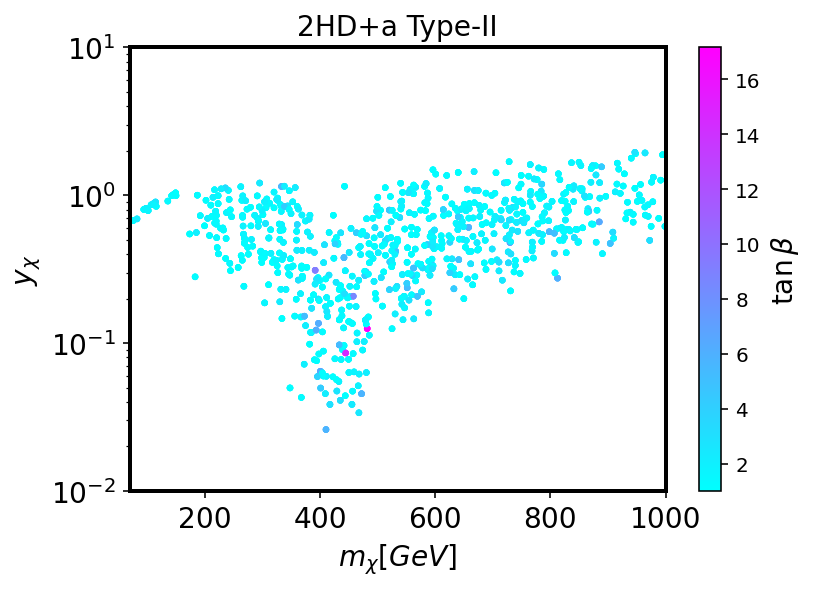}}
    \caption{Parameter points of the 2HD+a model satisfying the theoretical, collider, flavor, and loop-induced SI direct-detection constraints discussed in the previous sections, together with the requirement of reproducing the observed DM relic abundance within the adopted $3\sigma$ interval. The light-pseudoscalar mass is varied in the range $1~{\rm GeV}\leq m_a\leq10~{\rm GeV}$. The accepted points are shown in the $[m_\chi,y_\chi]$ plane, with the color scale indicating $\tan\beta$. The left and right panels correspond to the Type-I and Type-II Yukawa realizations, respectively.}
    \label{fig:plot_relic}
\end{figure}

The resulting parameter space is shown in Fig.~\ref{fig:plot_relic}. We display explicitly the Type-I and Type-II realizations, which are representative of the two qualitatively different quark-Yukawa structures considered throughout this work. For the relic-density phenomenology, Type-X and Type-Y closely follow Type-I and Type-II, respectively, over much of the parameter space, in particular when annihilation into quark or Higgs-sector final states dominates.
Figure~\ref{fig:plot_relic} shows that the relic-density requirement selects values of $y_\chi$ typically of order unity or smaller over most of the viable parameter space. This behavior reflects the fact that thermal annihilation can remain efficient even for moderate DM couplings, owing to the pseudoscalar-mediated annihilation into SM fermions and, whenever kinematically accessible, into additional Higgs-sector states. The distribution is broad because the annihilation rate depends not only on $y_\chi$, but also on $m_a$, the heavy-scalar spectrum, $\sin\theta$, $\tan\beta$, and the availability of the different final states.

\section{Combined results}
\label{sec:combined}

\begin{figure}
    \centering
    \subfloat{\includegraphics[width=0.45\linewidth]{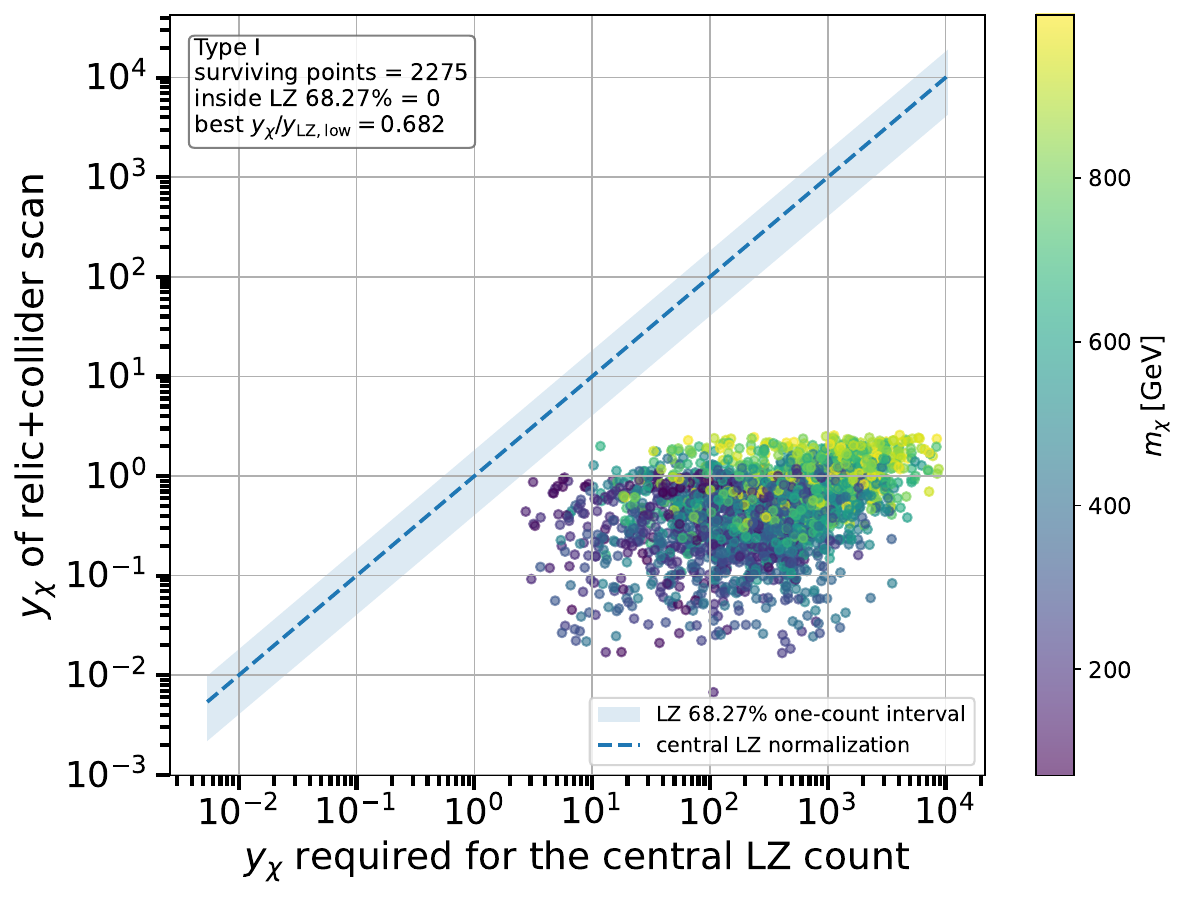}}
    \subfloat{\includegraphics[width=0.45\linewidth]{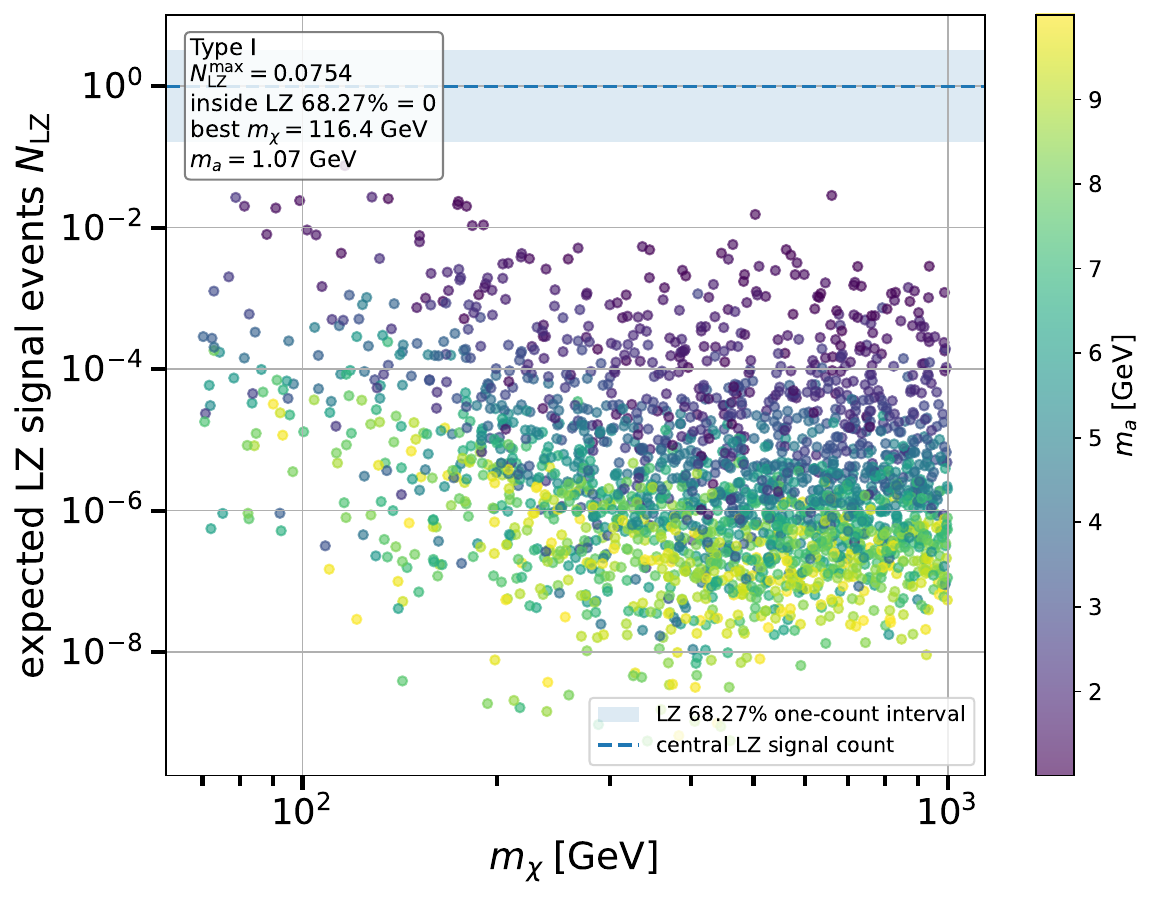}}\\
    \subfloat{\includegraphics[width=0.45\linewidth]{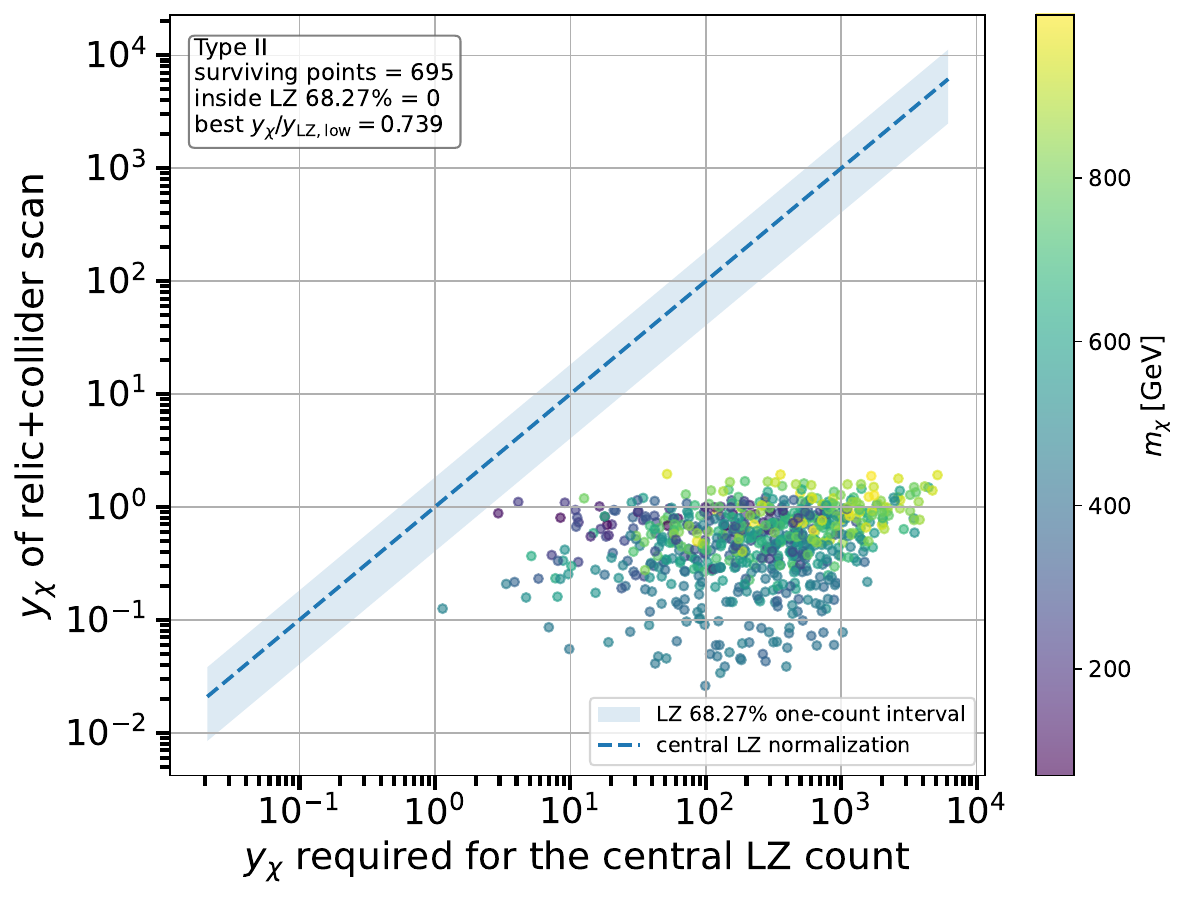}}
    \subfloat{\includegraphics[width=0.45\linewidth]{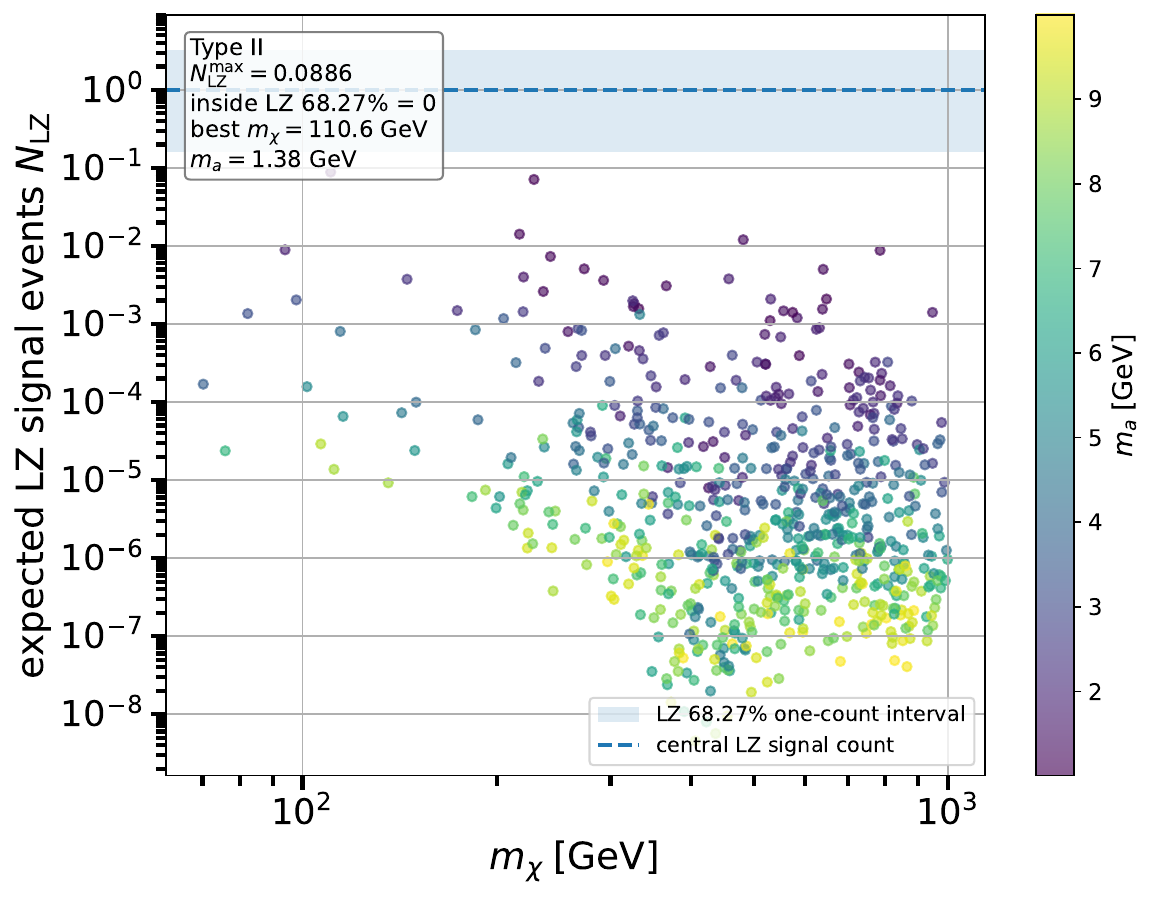}}
    \caption{Comparison between the parameter space satisfying the thermal relic-density requirement and the normalization required by the LZ candidate event. The upper and lower rows correspond to the Type-I and Type-II realizations, respectively. Left panels: value of $y_\chi$ selected by the relic-density and complementary constraints as a function of the coupling $y_\chi^{\rm LZ}$ required to reproduce the central LZ signal normalization for the same parameter point. The dashed diagonal corresponds to $y_\chi=y_\chi^{\rm LZ}$, while the shaded band denotes the 68.27\% one-count Poisson interval. The color scale indicates $m_\chi$. Right panels: expected number of LZ signal events $N_{\rm LZ}$ for the same accepted parameter points as a function of $m_\chi$, with the color scale indicating $M_a$. The dashed horizontal line denotes the central one-event normalization and the shaded band its 68.27\% Poisson interval. Analogous qualitative behavior is obtained for Type-X and Type-Y, which closely follow Type-I and Type-II, respectively, for the relic-density phenomenology.}
    \label{fig:LZ_relic_combined}
\end{figure}

The central question for the present analysis is whether the coupling selected by thermal freeze-out can simultaneously reproduce the LZ candidate event. We therefore compare directly, for every viable relic-density point, the value of $y_\chi$ obtained from the parameter scan with the coupling required to reproduce the LZ normalization. The result is shown in Fig.~\ref{fig:LZ_relic_combined}.

The left panels of Fig.~\ref{fig:LZ_relic_combined} provide the most direct comparison between the two requirements. The horizontal coordinate gives the coupling $y_\chi^{\rm LZ}$ that would be required to reproduce the central LZ normalization for each parameter configuration, while the vertical coordinate gives the actual value of $y_\chi$ selected by the parameter scan after imposing the relic-density and complementary constraints. Exact agreement with the central LZ normalization would therefore correspond to points lying on the dashed diagonal, while the shaded region indicates the broader coupling interval associated with the 68.27\% one-count Poisson range.

The surviving points lie systematically below the region preferred by the central LZ normalization. This shows that, within the conventional thermal freeze-out scenario considered here, the relic-density requirement generally favors smaller values of $y_\chi$ than those needed to reproduce one expected LZ event. The tension can be quantified relative to the lower edge of the 68.27\% one-count interval. For the Type-I scan, the point closest to the LZ-preferred region reaches approximately
\begin{equation}
\frac{y_\chi}{y_{\chi,{\rm LZ}}^{\rm low}}
\simeq 0.68,
\end{equation}
while for Type-II the corresponding ratio is approximately
\begin{equation}
\frac{y_\chi}{y_{\chi,{\rm LZ}}^{\rm low}}
\simeq 0.74.
\end{equation}
Thus, none of the accepted scan points enters the 68.27\% LZ interval, although the two requirements approach each other rather closely in coupling space.

The same conclusion is illustrated in a complementary way by the right panels of Fig.~\ref{fig:LZ_relic_combined}, where we show the expected number of LZ signal events for each viable relic-density point. The largest predicted rates are
\begin{equation}
N_{\rm LZ}^{\rm max}\simeq0.075
\end{equation}
for Type-I and
\begin{equation}
N_{\rm LZ}^{\rm max}\simeq0.089
\end{equation}
for Type-II. These values remain below the lower boundary of the 68.27\% one-event Poisson interval and are substantially below the central normalization $N_{\rm LZ}=1$.

It is important, however, to account for the large statistical uncertainty associated with a single observed event. For one event, the exact central 95.45\% Poisson interval for the underlying mean signal expectation is approximately
\begin{equation}
0.023 \lesssim N_{\rm LZ} \lesssim 5.68.
\label{eq:LZ_2sigma_poisson}
\end{equation}
Both maximal signal expectations found in our scan,
\begin{equation}
N_{\rm LZ}^{\rm max}\simeq0.075
\qquad {\rm and} \qquad
N_{\rm LZ}^{\rm max}\simeq0.089,
\end{equation}
therefore lie inside the $2\sigma$ Poisson interval. The combined collider, direct-detection, and relic-density analysis is consequently compatible with the LZ observation at the 95.45\% confidence level, despite the absence of overlap with the narrower 68.27\% interval.

The same statement can be expressed directly in terms of the DM coupling. Since the tree-level LZ event rate scales approximately as
\begin{equation}
N_{\rm LZ}\propto y_\chi^2,
\end{equation}
the lower boundary of the 95.45\% one-event interval corresponds to
\begin{equation}
\frac{y_\chi}{y_\chi^{\rm LZ}}
\simeq
\sqrt{0.023}
\simeq0.15,
\end{equation}
where $y_\chi^{\rm LZ}$ denotes the coupling required for the central normalization $N_{\rm LZ}=1$. By comparison, the maximal Type-I and Type-II rates correspond approximately to
\begin{equation}
\frac{y_\chi}{y_\chi^{\rm LZ}}
\simeq
\sqrt{0.075}
\simeq0.27
\end{equation}
and
\begin{equation}
\frac{y_\chi}{y_\chi^{\rm LZ}}
\simeq
\sqrt{0.089}
\simeq0.30,
\end{equation}
respectively. The best viable points therefore remain comfortably above the lower edge of the $2\sigma$ Poisson region.

The points that come closest to the LZ signal after imposing the collider, SI direct-detection, and relic-density constraints occupy a rather specific region of parameter space, characterized by a DM mass close to $100~{\rm GeV}$ and a very light pseudoscalar. In the Type-I scan, the largest predicted LZ rate is obtained for
\begin{equation}
m_\chi\simeq116.4~{\rm GeV},
\qquad
m_a\simeq1.07~{\rm GeV},
\qquad
\tan\beta\simeq1.53,
\end{equation}
with $|\sin\theta|\simeq0.24$, $M_A\simeq985~{\rm GeV}$, and $y_\chi\simeq0.87$. This point predicts $N_{\rm LZ}\simeq0.075$. For Type-II, the most favorable surviving configuration is found for
\begin{equation}
m_\chi\simeq110.6~{\rm GeV},
\qquad
m_a\simeq1.38~{\rm GeV},
\qquad
\tan\beta\simeq2.18,
\end{equation}
with $|\sin\theta|\simeq0.18$, $M_A\simeq917~{\rm GeV}$, and $y_\chi\simeq0.88$, yielding $N_{\rm LZ}\simeq0.089$. Both expected rates lie inside the 95.45\% Poisson interval associated with one observed event. The combined analysis therefore singles out a particularly interesting region with $m_\chi\simeq100$--$120~{\rm GeV}$, $m_a\simeq1$--$1.5~{\rm GeV}$, moderate pseudoscalar mixing, and relatively small $\tan\beta\simeq1.5$--$2.2$. This confirms that the preference for a very light mediator persists after imposing the complementary constraints, while showing that the most viable combined solutions do not occur in the large-$\tan\beta$ regime favored by the tree-level Type-II LZ normalization alone.

The comparison therefore reveals a mild tension at the level of the 68.27\% one-event interval between the standard thermal-relic interpretation and the coupling required to reproduce the central LZ normalization. Importantly, this tension does not persist at the $2\sigma$ Poisson level: viable points satisfying the collider, direct-detection, and thermal relic-density constraints predict signal rates that fall within the 95.45\% interval associated with one observed event. The 2HD+a interpretation is therefore not excluded by the combined analysis, although reproducing the central LZ normalization would require somewhat larger values of the DM coupling than those favored by thermal freeze-out.

\section{Conclusions}
\label{sec:conclusions}

In this paper we have investigated whether the recent high-recoil LZ excess can be interpreted in terms of elastic scattering of a fermionic DM candidate within the $2{\rm HDM}+a$ framework. The distinctive feature of this interpretation is that the tree-level pseudoscalar interaction matches onto the momentum- and SD non-relativistic operator $\mathcal{O}_6$. The resulting momentum dependence enhances the relative importance of high-energy nuclear recoils compared with conventional SI scattering and can therefore accommodate an event around $E_R\simeq248~{\rm keV}$. At the same time, a viable realization requires the loop-induced SI contribution associated with $\mathcal{O}_1$ to remain below the stringent conventional direct-detection limits.

Considering the LZ signal alone, the largest scattering rates are obtained for a relatively light pseudoscalar mediator, sizable singlet--doublet mixing, and sufficiently large DM Yukawa coupling. In Type-II and Type-Y, the enhancement of the coupling to down-type quarks at large $\tan\beta$ can substantially increase the direct-detection rate, whereas Type-I and Type-X favor values of $\tan\beta$ closer to unity. In all cases, the LZ normalization clearly favors the lightest pseudoscalar masses considered in our analysis.

This region is, however, subject to important complementary constraints. For $m_a<M_h/2$, searches for the exotic Higgs decay $h\rightarrow aa$ require a strong suppression of the trilinear coupling $\lambda_{haa}$, implying specific relations among the parameters of the scalar potential. Searches for light dimuon resonances and rare meson decays further constrain the low-$M_a$ region. These bounds are particularly restrictive for Type-I, Type-II, and Type-X. Type-Y is comparatively less affected by dimuon searches at large $\tan\beta$, since the enhanced coupling to down-type quarks is accompanied by a suppressed coupling to charged leptons, and therefore remains especially interesting from the collider perspective.

The picture becomes more restrictive once the thermal relic-density requirement is imposed simultaneously with the theoretical, collider, flavor, and SI direct-detection constraints. The values of $y_\chi$ preferred by thermal freeze-out are typically somewhat smaller than those required to reproduce the central LZ normalization. Consequently, we do not find surviving scan points inside the $68.27\%$ one-event Poisson interval.

The situation changes when the large statistical uncertainty associated with a single observed event is taken into account. Using the central $95.45\%$ Poisson interval, we find surviving parameter points that are compatible with the LZ observation. The most interesting region is characterized approximately by
\begin{equation}
m_\chi\sim100\text{--}200~{\rm GeV},
\qquad
m_a\sim1\text{--}2~{\rm GeV},
\qquad
\tan\beta\sim1\text{--}2,
\end{equation}
together with a DM Yukawa coupling of order unity and moderate pseudoscalar mixing, typically $|\sin\theta|\sim0.1$--$0.3$. Representative Type-I and Type-II solutions are found close to $m_\chi\simeq110~{\rm GeV}$ and $m_a\simeq1~{\rm GeV}$. Thus, the combined analysis does not reproduce the central LZ normalization, but it remains compatible with the observed event at the $2\sigma$ Poisson level.

An important consequence of the combined analysis is that the preferred parameter region differs from the one obtained by maximizing the LZ scattering rate alone. In particular, once the complementary constraints and the relic abundance are imposed, the surviving solutions favor moderate values of $\tan\beta$ and of the pseudoscalar mixing angle rather than the very large $\tan\beta$ or nearly maximal mixing that can enhance the unconstrained tree-level rate.

We therefore conclude that elastic pseudoscalar-mediated scattering in the $2{\rm HDM}+a$ model provides a viable phenomenological mechanism for producing a high-energy recoil of the type observed by LZ. A simultaneous explanation of the LZ event and of the thermal DM abundance is somewhat disfavored at the $1\sigma$ level, but remains possible at $2\sigma$ in a restricted region characterized by a light pseudoscalar, DM masses around the electroweak scale, and couplings of order unity. A more complete global analysis, particularly of the Type-Y realization and of the low-mass collider and flavor constraints, will be important for determining whether this residual region remains viable.

\section{Acknowledgments}

MDM acknowledges support from the research grant {\sl TAsP (Theoretical Astroparticle Physics)} funded by Istituto Nazionale di Fisica Nucleare (INFN).
AD was supported by the Spanish grant PID2021-128396NB-I00. 
FSQ was supported by Simons Foundation (Award Number:1023171-RC), FAPESP Grant 2018/25225-9, 2021/01089-1, 2023/01197-4, ICTP-SAIFR FAPESP Grants 2021/14335-0, CNPq Grants 307130/2021-5, and ANID-Millennium Science Initiative Program ICN2019\_044, and FINEP under the project 213/2024.

\bibliographystyle{utphys}
\bibliography{main}
\end{document}